\documentclass[12pt,a4paper]{article}
\usepackage{graphicx}
\usepackage{bm}

\begin{document}

\parskip=0mm
\parindent=5mm

\pagestyle{plain}

\noindent{\Large {\bf Observations and modelling of ion cyclotron}}

\vskip 0.2cm

\noindent{\Large {\bf emission observed in JET plasmas using a}} 

\vskip 0.2cm

\noindent{\Large {\bf sub-harmonic arc detection system during ion}}

\vskip 0.2cm

\noindent{\Large {\bf cyclotron resonance heating}}

\vskip 0.5cm

\noindent{\bf K.G. McClements$^1$, A. Brisset$^2$, B. Chapman$^3$, S.C. Chapman$^3$, R.O. Dendy$^{1,3}$, P. Jacquet$^1$, V.G. Kiptily$^1$, M. Mantsinen$^{4,5}$, B.C.G. Reman$^3$ and JET Contributors\footnote{See the author list of  ``X. Litaudon {\it et al} 2017 {\it Nucl. Fusion} {\bf 57} 102001''}}

\vskip 0.5cm

\noindent{EUROfusion Consortium, JET, Culham Science Centre, Abingdon, OX14 3DB, UK}

\noindent{$^1$ CCFE, Culham Science Centre, Abingdon, Oxfordshire, OX14 3DB, UK}

\noindent{$^2$ Universit\'e Paris-Sud, 91405 Orsay cedex, France}

\noindent{$^3$ Department of Physics, University of Warwick, Coventry CV4 7AL, UK}

\noindent{$^4$ Barcelona Supercomputing Center, Barcelona, Spain}

\noindent{$^5$ ICREA, Passeig de Llu\'is Companys 23, 08010 Barcelona, Spain}

\vskip 0.5cm

\noindent E-mail: Ken.McClements@ukaea.uk

\vskip 0.5cm

\section*{Abstract}

\noindent Measurements are reported of electromagnetic emission close to the cyclotron frequency of energetic ions in JET plasmas heated by waves in the ion cylotron range of frequencies (ICRF). Hydrogen was the majority ion species in all of these plasmas. The measurements were obtained using a sub-harmonic arc detection (SHAD) system in the transmission lines of one of the ICRF antennas. The measured ion cyclotron emission (ICE) spectra were strongly filtered by the antenna system, and typically contained sub-structure, consisting of sets of peaks with a separation of a few kHz, suggesting the excitation of compressional Alfv\'en eigenmodes (CAEs) closely spaced in frequency. In most cases the energetic ions can be clearly identified as ICRF wave-accelerated $^3$He minority ions, although in two pulses the emission may have been produced by energetic $^4$He ions, originating from third harmonic ICRF wave acceleration. It is proposed that the emission close to the $^3$He cyclotron frequency was produced by energetic ions of this species undergoing drift orbit excursions to the outer midplane plasma edge. Particle-in-cell and hybrid (kinetic ion, fluid electron) simulations using plasma parameters corresponding to edge plasma conditions in these JET pulses, and energetic particle parameters inferred from the cyclotron resonance location, indicate strong excitation of waves at multiple $^3$He cyclotron harmonics, including the fundamental, which is identified with the observed emission. These results underline the potential importance of ICE measurements as a method of studying confined fast particles that are strongly suprathermal but have insufficient energies or are not present in sufficient numbers to excite detectable levels of $\gamma$-ray emission or other collective instabilities.        

\section{Introduction}

\noindent Spontaneous emission of electromagnetic radiation in the ion cyclotron range of frequencies, usually referred to as ion cyclotron emission (ICE), is a frequently-observed property of tokamak plasmas containing suprathermal ion populations \cite{TFR1978,Cottrell1993,Ichimura2008,Heidbrink2011}. Links between such emissions and suprathermal ions have also been identified in the Large Helical Device (LHD) stellarator \cite{Saito2013} and in the Earth's radiation belts \cite{McClements1993,McClements1994,Posch2015}, indicating that the phenomenon is not confined to one specific magnetic configuration. ICE measurements, in combination with appropriate modelling, can be used to obtain diagnostic information on the behaviour of confined fast ions, including charged fusion products, ions accelerated by waves in the ion cyclotron range of frequencies (ICRF), and, in some devices, beam-injected ions. It has been proposed that ICE could be used to study fusion $\alpha$-particles in the burning plasma ITER device \cite{McClements2015}. Various types of ICE detector have been used \cite{McClements2015}, including detectors that are incorporated into ion cyclotron resonance heating (ICRH) antenna systems: this approach has been used successfully in the ASDEX Upgrade \cite{Dinca2011} and JET \cite{Jacquet2011} tokamaks. With the use of suitable filters to suppress signals corresponding to the launched ICRH frequencies, ICE detection is possible in this case during radio-frequency heating, thereby making it possible to detect emission excited by ICRH fast ions. Another important advantage of this technique is that it does not require the installation of any additional hardware inside the vacuum vessel \cite{Jacquet2011}. Dedicated probes have also been used to detect ICE driven by ICRH fast ions in both ASDEX Upgrade \cite{Dinca2011} and JET \cite{Cottrell2000}. 

First principles-based modelling of ICE in full toroidal geometry is challenging, since the emission frequencies are too high for the wave-particle interaction to be represented accurately in terms of a guiding centre approximation for the energetic ion orbits. However at least some of the features of ICE can be captured by a first principles, fully nonlinear, numerical solution of the Maxwell-Lorentz system of equations which evolves hundreds of millions of electrons and ions, including an unstable energetic ion population, together with the self-consistent electric and magnetic fields, in a locally uniform equilibrium plasma. One-dimensional codes that co-evolve the fields and the full orbits of the ions, while employing either a kinetic or a fluid description of the electrons, have been used to model the nonlinear evolution of ICE excited by fusion $\alpha$-particles in JET \cite{Cook2013,Carbajal2017}, and of ICE observed during edge localised modes in the medium-sized tokamak KSTAR \cite{Thatipamula2016,Chapman2017}. Notably, this approach has reproduced an observed linear scaling of fusion product-driven ICE intensity with neutron flux \cite{Carbajal2017}. Linear modelling of fundamental and second harmonic ICE driven by $^3$He fusion products in the large tokamak JT-60U has yielded estimates of the frequencies and toroidal mode numbers of the most rapidly growing modes which are in good agreement with measured values for ICE in this device \cite{Sumida2017}. In this case the growth rates were obtained using a model fast ion distribution in the ICE source region (the low field side plasma edge) which was based on calculated guiding centre orbits of $^3$He fusion products arising from beam-thermal reactions. These orbit calculations showed that $^3$He nuclei reaching the low field side edge had a narrow range of pitch angles and a non-monotonic distribution of energies, peaking at a value close to the birth energy (0.82$\,$MeV).                           

In this paper we present a more detailed analysis of the ICE measurements from JET first reported in Ref. \cite{Jacquet2011}, and also modelling of these measurements carried out using the {\tt EPOCH} full particle-in-cell (PIC) code \cite{Arber2015} and a hybrid code which treats ions kinetically and the electrons as a fluid \cite{Carbajal2014}. Following a brief description of the ICE detection system in JET ICRH antennas (section 2), measurements of ICE in a series of hydrogen pulses with ICRH are discussed in section 3. In most of these pulses the species of fast ions driving the emission can be clearly identified as $^3$He, but in two pulses it is more likely to have been $^4$He. In section 4 we present calculations of the $^3$He distribution in a particular pulse, which indicate the presence of multi-MeV fast ions with orbits that traverse the low field side plasma edge. Results of one-dimensional PIC and hybrid simulations, showing excitation of emission at the $^3$He cyclotron frequency and its harmonics due to the presence of such ions, are then presented. The successes and limitations of the one-dimensional PIC and hybrid techniques as ICE modelling tools, and the possible exploitation of ICE as a fast particle diagnostic in future experiments, are discussed in section 5.     
 
\section{Sub-harmonic arc detection system in JET}

The measurements of ICE reported by Jacquet and co-workers \cite{Jacquet2011} were obtained using a sub-harmonic arc detection (SHAD) system in the JET ICRH antennas (see figure 1). The SHAD method is based on the detection of signals in the (2-20)$\,$MHz band, which in principle represent the signature of arc events in the ICRH antenna and transmission line system. However, signals emitted by the plasma were also detected. The system was installed in the so-called A2 antenna D \cite{Graham2012}. A low pass filter was applied to suppress frequencies above 25$\,$MHz. SHAD data acquisition was performed using a digital oscilloscope with a sampling rate of 125 Msample$\,$s$^{-1}$. The total acquisition time window was typically 128$\,$ms. Acquisition was triggered when the signal intensity exceeded a certain threshold. The frequency content of the recorded signals was further determined using numerical fast Fourier transform analysis.

The radio-frequency (RF) signals are sampled in the ICRF antenna transmission lines, which act as RF resonators. The frequency response of these resonators has to be taken into account when analysing the signals recorded by the SHAD system. The form of the response in frequency depends on the exact antenna configuration (in particular whether C and D are fed independently or through the external conjugate-T scheme), and on the length of the transmission line matching elements, which are adjusted according to the ICRH frequency used and the antenna-plasma coupling conditions. Figure 2 shows the calculated frequency response for signals entering the transmission lines of antenna D (fed independently) when tuned for ICRH operation at 51.53$\,$MHz, which was used in pulse \#79363 described in section 3.

\section{ICE 	measurements using the SHAD system}

\noindent Hydrogen was the dominant ion species in all of the JET pulses discussed in this paper: $^3$He ions were present as a minority species with a concentration of up to about 12\%. The minority concentration was determined by comparing the intensity of a $^3$He emission line to that of D$\alpha$ emission in the outer divertor region. This is a plasma edge measurement, but the concentration estimated using this method is consistent with a comparison between experimental and modelled radio-frequency heating efficiencies performed in these pulses \cite{Lerche2012}, and therefore the divertor measurements are believed to provide a reasonable estimate of the $^3$He concentration in the plasma core. The minority ions were accelerated by ICRF waves via the second harmonic resonance. The detection of $\gamma$-ray emission and fast ion losses indicated the presence in these pulses of $^3$He ions with energies in the MeV range \cite{Kiptily2012}. Trace quantities of deuterium (D) and $^4$He ions were also present, principally due to the use of deuterium neutral beam injection and wall recycling \cite{Kiptily2012}. 

Table 1 lists some of the relevant parameters for ten shots in which ICE was observed, including the vacuum toroidal field at a reference major radius of 2.96$\,$m ($B_{\rm T}$), and the ICRF power ($P_{\rm RF}$) and neutral beam power ($P_{\rm NBI}$) at the time $t_{\rm ICE}$ of ICE detection. In this table $R_{\rm He}$ is an estimate of the major radius at which the measured ICE frequency $\nu_{\rm ICE}$ was equal to the $^3$He cyclotron frequency, based on the value of $B_{\rm T}$. In every case $R_{\rm He}$ is comparable to or somewhat higher than the major radius $R_{\rm LFS}$ of the low field side midplane plasma edge (typically $R_{\rm LFS} \simeq 3.9\,$m for the pulses and times considered here). The fact that ICE was detected at the same frequency (19.2$\,$MHz) in seven pulses with ICRF wave-accelerated minority $^3$He ions and the same toroidal field (2.62$\,$T) illustrates the  reproducibility of the phenomenon. JET had a carbon wall at the time of these pulses, which were in the low confinement mode of operation (``L-mode''). Each of the pulses had a plasma current of 1.4$\,$MA, except for pulses \#79364 and \#79365 which had a plasma current of 2.0$\,$MA, and the central electron density at the time of ICE detection ranged from $2.3\times 10^{19}$m$^{-3}$ to $3.4\times 10^{19}$m$^{-3}$.

\vskip 0.4cm

\noindent {\bf Table 1.} Key parameters of JET hydrogen pulses in which ICE was detected using SHAD system.  

\vskip 0.4cm

\centerline{\begin{tabular}{ccccccc} \hline
Pulse & $B_{\rm T}\,$(T) & $P_{\rm RF}\,$(MW) & $P_{\rm NBI}\,$(MW) & $t_{\rm ICE}\,$(s) & $\nu_{\rm ICE}\,$(MHz) & $R_{\rm He}\,$(m) \\ \hline
79355 & 2.62 & 1.4 & 1.3 & 8.1 & 18.0 & 4.3\\ 
79356 & 2.62 & 1.7 & 6.6 & 4.6 & 19.2 & 4.1\\ 
79357 & 2.62 & 3.0 & 5.3 & 4.6 & 19.2 & 4.1\\ 
79358 & 2.62 & 1.3 & 5.3 & 4.0 & 19.2 & 4.1\\ 
79359 & 2.62 & 3.0 & 2.7 & 4.4 & 19.2 & 4.1\\ 
79361 & 2.62 & 4.5 & 2.6 & 4.4 & 19.2 & 4.1\\ 
79362 & 2.62 & 5.4 & 8.0 & 4.7 & 19.2 & 4.1\\ 
79363 & 2.62 & 5.0 & 1.3 & 6.9 & 19.2 & 4.1\\ 
79364 & 2.25 & 3.0 & 3.2 & 17.2 & 15.0 & 4.5\\ 
79365 & 2.25 & 5.7 & 7.2 & 16.5 & 14.8 & 4.5\\ \hline
\end{tabular}}

\vskip 0.4cm

Figure 3 shows the ICRH resonance layers for $^3$He and $^4$He/D in two of the pulses, (a) \#79363 and (b) \#79365, overlaid on equilibrium flux surfaces at the time of ICE excitation (the equilibria discussed in this paper were computed from current and coil measurements using the EFIT code \cite{Lao1985}). These pulses had different toroidal fields, but the ICRF frequencies were also different, with the result that the resonance layers almost coincided in the two cases. The key point to note here is that while the second harmonic $^3$He resonance lay close to the magnetic axis, as required for efficient minority heating, the second and third harmonic $^4$He/D resonances also lay well within the plasma, creating the possibility that ions of these species could have been accelerated to high energies.      

In several pulses the SHAD system was triggered by the detection of ICE during or shortly after the ramp-up in $P_{\rm RF}$, suggesting a strong correlation between the $^3$He fast ion population and the ion cyclotron emission. In one pulse, \#79358, ICE was detected almost immediately after the start of the ICRH ramp. The $^3$He concentration was particularly low in this case (approximately 1.5\%), suggesting that higher ICRF power per resonant particle could have resulted in faster formation of a high-energy tail in the $^3$He distribution function. In some of the pulses the trigger for ICE was less obvious, although the essential features of the emission were similar. In pulse \#79363 ICE was strongly excited during a period when the ICRH power alternated between 5.4 MW and 0.5 MW, with a frequency of 4 Hz. Detection was triggered approximately 400 ms after $P_{\rm NBI}$ power was stepped down from 6.6 MW to 1.3 MW, and, as shown in the lower frame of figure 4, immediately after the disappearance of a coherent magnetic fluctuation initially excited at a frequency of around 10 kHz with a toroidal mode number $n=2$. Charge exchange and electron cyclotron emission data, combined with equilibrium reconstructions based on magnetic coil data, indicate that the mode was located just inside a flux surface with safety factor $q=2$. This suggests that it was a neoclassical tearing mode  with dominant poloidal mode number $m=3$, i.e. a 3/2 mode. No sawtooth oscillations occurred during this period, suggesting that $q$ in the plasma core remained above unity. The drop in mode frequency apparent in figure 4 coincided with a slowing down of the toroidal plasma rotation, from about 120 km s$^{-1}$ to less than 60 km s$^{-1}$ in the plasma core. The upper frame of figure 4 shows that ICE was first detected as soon as the ICRH power was raised to its flat-top value of 5.4 MW after the disappearance of the mode. During the period of $n=2$ mode excitation, high levels of fast particle losses were found to be correlated with the peaks in ICRH power, during a period in which the neutron rate remained almost constant; for this reason the losses were believed to be mainly $^3$He ICRH fast ions. A much lower loss rate was recorded during the ICRH peak that coincided with ICE excitation (see figure 34 in Ref. \cite{Kiptily2012}). At this time, no coherent low frequency MHD modes were detected.     

Figure 5a shows the temporal evolution of the ICE intensity in pulse \#79363 during the 60$\,$ms after the initial detection at 6.9$\,$s (the ICRH power remained close to its peak value throughout this period), while figure 5b shows an example of the recorded spectrum of this emission (at $t \simeq 6.94\,$s). The intensity plot was obtained by integrating the emission over the frequency interval 18.7$\,$MHz to 19.7$\,$MHz and time windows of 100$\,\mu$s duration, and the spectrum was generated using data acquired over a 700$\,\mu$s time interval. It can be seen that the recorded intensity varies significantly between successive time windows. In figure 5b the emission peaks at 19.25 MHz, which is close to but slightly lower than the $^3$He cyclotron frequency $\Omega_{^3{\rm He}}/2\pi = 20.7\,$MHz on the low field side plasma edge at this time ($R \simeq 3.86\,$m). This was also found to be the case for $^3$He ICE in JT-60U \cite{Sumida2017}. The use of a linear scale, rather than the more conventional logarithmic scale, makes it clear that there is sub-structure within the envelope of emission. This sub-structure appears to be a true feature of the emission rather than an instrumental effect or noise: the frequency spacing of the individual sub-peaks (a few kHz) is much smaller than the frequency separation of the peaks in the response function of the antenna transmission lines (about 1.4$\,$MHz: see figure 2), and smaller than the full width at half maximum of these response function peaks (about 80$\,$kHz). Moreover the frequencies of the most intense sub-structure peaks in ICE spectra persist in successive time slices for a given pulse, although the relative intensities of these peaks change rapidly in time.

In the two pulses immediately following \#79363, emission of a similar character was detected at significantly lower frequencies, around 15$\,$MHz. In these pulses the toroidal field was slightly lower than in \#79363, such that the $^3$He cyclotron frequency at the outer midplane plasma edge $R=R_{\rm LFS}$ at the time of ICE detection was approximately 17.4$\,$MHz, while $R_{\rm He}$ was significantly higher than $R_{\rm LFS}$ (see Table 1). The $^4$He/D cyclotron frequency at $R=R_{\rm LFS}$ was approximately 13.1$\,$MHz (as noted at the start of this section, trace quantities of both species were present in the plasma). The ICRH frequencies were reduced in proportion to the toroidal field, so that the second harmonic $^3$He resonance was in approximately the same location as in pulse \#79363 ($R \simeq 3.15\,$m). In the case of pulse \#79365, ICE detection was triggered at $t \simeq 16.5\,$s, at the end of the ramp-up in ICRH power. Unlike pulse \#79363, the onset of ICE in this case was not correlated with a change in low frequency MHD activity. Figure 6a shows the temporal evolution of the recorded ICE intensity in the 60$\,$ms following the trigger, while figure 6b shows the spectrum at $t =\simeq 16.51\,$s). Sub-structure peaks in the spectrum are again apparent, and again these peaks appear in successive time slices.

The measured ICE frequency in pulse \#79365 lies between the $^3$He and $^4$He/D cyclotron frequency at the outer plasma edge, although it is somewhat closer to the latter. It is thus not clear from the SHAD data alone which fast ion species is driving the ICE. However some information on the composition of the fast ion population in the edge plasma during this pulse can be obtained from measurements of fast ion losses obtained using a scintillator probe (SP) \cite{Kiptily2012,Kiptily2009}. Relatively low levels of fast ion losses were recorded until $t \simeq 20.0\,$s (approximately 3.5$\,$s after the start of ICE detection), when there was a five-fold increase in the total loss rate. At this time the ICRH power was increased from 3$\,$MW to 5$\,$MW; an additional, more modest rise in fast particle losses was recorded when the neutral beam power was increased from 4$\,$MW to 10$\,$MW at 60.5$\,$s. Figure 7 shows the footprint of losses on the SP at this time. The red lines indicate the local pitch angles of protons, $^3$He ions and $^4$He/D ions accelerated by the applied ICRF waves at, respectively, the fundamental, second harmonic and third harmonic resonances. This footprint shows very clearly that the pitch angles of the lost fast ions are those expected of deuterons and $^4$He ions. The gyro-radii of the lost particles recorded in figure 7 suggest the presence of either deuterons with energies of around 1.1$\,$MeV or $^4$He ions with energies of around 2.2$\,$MeV. However $\gamma$-rays produced in the reactions $^{12}$C(d,p$\gamma$)$^{13}$C and $^9$Be($^4$He,n$\gamma$)$^{12}$C were not observed. It is possible that the losses shown in figure 7 could be due to singly-ionised $^4$He ions with energies around 0.55$\,$MeV, arising from the slowing down and charge-exchange of doubly-ionised 1.3$\,$MeV $^4$He ions at the probe entrance, where a 1$\,\mu$m Au foil was used to attenuate the flux of neutral beam ions and heavy impurity ions \cite{Kiptily2012}. The energies of such ions would be too low for them to produce $\gamma$-ray emission, but they could still excite ICE, provided that the local $^4$He velocity distribution satisfied the relevant instability criteria, discussed in the next section. 

The absence of the $\gamma$-ray lines mentioned above, combined with the SP footprint shown in figure 7 and the fact that the ICE frequency  in pulse \#79365 was somewhat closer to the $^4$He/D cyclotron frequency than the $^3$He cyclotron frequency at the low field side plasma edge, suggests that the cyclotron emission in this case may have been driven by trace ICRF wave-accelerated $^4$He ions. This is not certain, since the fast ions that are lost at a particular poloidal location (and recorded in SP footprints) are not necessarily of the same species as fast ions that are marginally confined in the plasma (and thus capable in principle of exciting ICE). Moreover it should be noted that there was a delay of 3.5$\,$s between the time of ICE detection and the time of strong fast ion losses in this pulse. Nevertheless it is most likely that the ICE in this pulse was driven by $^4$He fast ions. The absence of $\gamma$-ray emission makes it unlikely that significant numbers of $^4$He nuclei resulting from the deuterium-$^3$He fusion reaction were present, since these have birth energies of around 3.7$\,$MeV, and so it appears that a possible source of $^4$He fast ions was third harmonic ICRH. As shown in figure 3b, the third harmonic $^4$He resonance was located well within the plasma, on the low field side of the magnetic axis, suggesting that this explanation of the ICE detected in pulse \#79365 is credible.  

\section{Interpretation and modelling}

\noindent It is generally accepted that ICE in tokamak plasmas results from the excitation of waves on the fast Alfv\'en/ion Bernstein branches via the magnetoacoustic cyclotron instability (MCI), driven by energetic ions that are characterised by some form of population inversion, for example a velocity distribution perpendicular to the magnetic field that is not monotonic decreasing \cite{Cottrell1993,Belikov1976,Dendy1994}. Such distributions can arise in the low field side plasma edge of a tokamak plasma due to large drift orbit excursions of trapped energetic ions originating from the plasma core region (due to either fusion products or ICRH). Since the magnitude of the drift orbit excursion increases with particle energy and is pitch angle-dependent, it is possible for the local energetic ion velocity distribution in the low field side plasma edge to be highly singular, thereby providing strong drive for the MCI. 

\subsection{Fast ion distribution}

Energetic ion distributions resulting from ICRH can be modelled for JET pulses using the Fokker-Planck {\tt PION} code, which employs a quasi-linear operator to describe the absorption by resonant particles of radio-frequency wave power and computes also the effects of collisions on these particles \cite{Eriksson1993}. Drift orbit excursions of the particles are taken into account. Figure 8 shows profiles obtained using {\tt PION} for the total and parallel energy density (top) and number density (bottom) of energetic $^3$He ions in JET pulse \#79363, $t = 6.9\,$s, i.e. the time at which ICE was detected in this pulse. The radial coordinate $s$ in these profiles is defined as the square root of the poloidal flux $\psi_{\rm p}$ (set to zero at the magnetic axis) normalised to its value $\psi_{\rm pa}$ at the plasma edge. One of the input parameters for this simulation is the $^3$He concentration $\eta$ (the ratio of total $^3$He density to electron density). In the case of figure 8 the concentration was assumed to be 7.1\%, which is close to the best estimate of the measured value for this pulse and time. Two important points to note from this figure are, first, that the fast ion energy is almost entirely represented by motion perpendicular to the magnetic field, as is usually the case for ICRF-wave accelerated fast ions, and, second, that the fast ion number density $n_f$ falls off more rapidly than the energy density $e_f$ in the outer regions of the plasma ($s > 0.4$). This is a consequence of the energy-dependence of fast ion drift orbit excursions noted above, and illustrates the point that the local fast ion distribution close to the plasma edge is likely to be non-monotonic decreasing in energy.    

Figure 9 shows the $^3$He energy distribution obtained using {\tt PION} close to the peak of the profiles shown in figure 8 ($s \simeq 0.2$)  for two different values of $\eta$. It can be seen that the distribution extends to energies of about 10$\,$MeV, and is insensitive to $\eta$. The weakness of this dependence on minority ion concentration is due to the fact that the damping of wave power at the second harmonic ICRF resonance is a finite Larmor radius effect, which effectively determines the details of the resulting fast ion distribution function \cite{Mantsinen1999, Salmi2006}. Based on modelling of $\gamma$-ray emission detected in pulse \#79363 and similar pulses, Kiptily and co-workers \cite{Kiptily2012} inferred that the $^3$He fast ions in these pulses had energies below approximately 2$\,$MeV, suggesting that in fact there were few if any confined $^3$He ions in this pulse with energies in the range 2-10$\,$MeV. It should be noted that non-classical transport of energetic ions resulting from MHD activity is not modelled by {\tt PION}. As noted in the previous section, a large amplitude $n=2$ mode, correlated with a high level of fast ion losses, had disappeared by the time of ICE excitation in pulse \#79363. This suggests that the $^3$He fast ions were being transported radially by the $n=2$ mode, and that the disappearance of this mode led to a restoration of classical fast ion confinement. When this occurred, the fast ion population in the outer midplane edge (the most probable region of MCI excitation) is likely to have been dominated by $^3$He nuclei with the highest energies, since these would have undergone the largest orbit excursions. 

To illustrate the magnitude of these excursions for $^3$He fast ions in pulse \#79363, we have calculated the full orbits of ions initially with pitch angles close to 90$^{\circ}$ at the major radius corresponding to the cyclotron resonance, $R_{\rm res} \simeq 3.1\,$m. Figure 10 shows the orbits of ions with energies of (a) 2$\,$MeV and (b) 10$\,$MeV, in line with the range of $^3$He energies indicated by {\tt PION} and $\gamma$-ray modelling. The initial vertical positions were chosen to be such that the orbits intersected the outer midplane edge. The 10$\,$MeV ion was initially close to the peak of the profiles shown in figure 8 ($s = 0.16$), whereas the 2$\,$MeV ion started its orbit at a point where the fast ion density was relatively low ($s = 0.66$). On the other hand, figure 9 suggests that 2$\,$MeV ions were much more numerous than 10$\,$MeV ions (even without taking into account the evidence from $\gamma$-ray emission that few if any confined 10$\,$MeV ions were present). The approximate location of the $q=3/2$ surface at the time of ICE excitation, estimated using EFIT, is shown as green curves on figure 10. This lies close to the plasma centre, lending weight to the suggestion above that the large amplitude mode shown in figure 4 could have affected the radial profile of core-localised ICRF fast ions.             

\subsection{Particle-in-cell simulations}

We have used {\tt EPOCH} to model the excitation and nonlinear evolution of the MCI, using bulk plasma parameters comparable to those in the plasma edge at the time of ICE detection in JET pulse \#79363, and fast ion parameters that are compatible with the calculations presented in the previous subsection. Figure 11 shows Thomson scattering measurements of electron density (left) and electron temperature (right) profiles for this pulse and time (the measurements have been averaged over a 400$\,$ms time interval to reduce noise). As noted previously, this was an L-mode plasma, in which the density did not have a steep gradient in the edge region, but the range of densities shown in figure 11 nevertheless means that the value chosen for this parameter in the {\tt EPOCH} simulations is somewhat arbitrary. In all of the simulations presented in this paper, we chose the initial plasma density to be $10^{19}\,$m$^{-3}$. The magnetic field at the low field side plasma edge $B_{\rm LFS}$ was approximately 2.0T in this pulse. For the case of a pure hydrogen plasma, these figures indicate an Alfv\'en speed (and hence the propagation speed of waves excited via the MCI) of $c_A \simeq 1.4 \times 10^7\,$ms$^{-1}$. We also adopt a value of 100$\,$eV for the initial temperature of both the electrons and the bulk ions. The fast ions were assumed to have an initial velocity distribution $f$ consisting of a monoenergetic ring-beam of the form
$$ f = {n_f\over 2\pi v_{\perp 0}}\delta(v_{\parallel}-v_{\parallel 0})\delta(v_{\perp}-v_{\perp 0}), \eqno (1) $$    
where, as before, $n_f$ is the total fast ion density, $v_{\parallel}$ and $v_{\perp}$ denote velocity components parallel and perpendicular to the magnetic field, and $v_{\parallel 0}$, $v_{\perp 0}$  are constants. This choice of initial distribution is motivated by the fact that the local $^3$He distribution is likely to have been dominated by high energy ions with a narrow range of pitch angles, as noted above. Here we make the conservative assumption that the maximum energy of confined $^3$He fast ions was 2$\,$MeV, and make the reasonable further assumption that the ions were accelerated to this energy perpendicular to the magnetic field in a region of the plasma with magnetic field $B=B_{\rm res}$ satisfying the second harmonic cyclotron resonance condition, i.e. $2\pi\nu_{\rm RF} = 2 \times 2eB_{\rm res}/3m_p$ where $\nu_{\rm RF}  = 51\,$MHz is the ICRH frequency and $e$, $m_p$ are proton charge and mass. We deduce from magnetic moment conservation that the most energetic $^3$He ions in the outer midplane plasma edge had velocity components parallel and perpendicular to the local magnetic field given by
$$ v_{\parallel 0} = v_{\rm max}\left(1-{B_{\rm LFS}\over B_{\rm res}}\right)^{1/2}, \;\;\;\;\;\;   v_{\perp 0} = v_{\rm max}\left({B_{\rm LFS}\over B_{\rm res}}\right)^{1/2}, \eqno (2) $$        
where $v_{\rm max}$ is the speed of a 2.0 MeV $^3$He ion. Evaluating these expressions, we deduce that appropriate values for the parameters $v_{\parallel 0}$ and $v_{\perp 0}$ in equation (1) are, respectively, $5 \times 10^6\,$ms$^{-1}$ and $10^7$ms$^{-1}$.

Simulations in one space dimension and three velocity dimensions (1D3V) were performed using {\tt EPOCH} for a uniform equilibrium plasma with hydrogen bulk ions and energetic $^3$He minority ions with an assumed local concentration of $10^{-3}$: since we are assuming that the fast $^3$He population in the plasma edge is dominated by MeV energy ions with large excursion orbits, the fast ion fraction is necessarily much smaller than the total $^3$He concentration in this pulse ($\sim 7\%$). We will discuss later the effects on the MCI of varying the fast ion concentration. The other initial parameters of each simulation were those listed above. The equilibrium magnetic field $\bm{B}_0$ was either taken to be exactly orthogonal to the space direction (i.e. the wavevector direction of any modes excited) or offset from this by one degree. In all cases the simulation box consisted of 70,000 cells (with 900 particles per cell), thereby making it possible to represent a very wide range of wavenumbers. Periodic boundary conditions were used. The duration of each simulation was ten $^3$He cyclotron periods; since this is much shorter than any collision time under these conditions, there was no need to take collisional processes into account.  

In each simulation the MCI was found to be excited at multiple harmonics of the $^3$He cyclotron frequency, leading to a relaxation of the energetic ion distribution, and a transfer of energy from the energetic ions to the other plasma species and the fields. This is shown in figure 12 for wave propagation angles $\theta$ of (a) 90$^{\circ}$ and (b) 89$^{\circ}$. When  $\theta = 90^{\circ}$ the energetic ions (cyan curve) lose nearly 10\% of their initial energy due to the MCI, but regain most of this in the nonlinear phase of the instability. The net energy lost from the energetic ions is transferred mainly to the bulk ions (red curve), with a smaller fraction being channelled into fluctuating $E_x$ and $B_z$ fields (blue and green curves respectively), and a smaller still fraction being transferred to electrons (black curve). The $x$-direction here is the single space direction of the simulation and the equilibrium magnetic field lies in the $z$-direction; $E_x$ thus represents the electrostatic (ion Bernstein) component of the fluctuations, while $B_z$ represents the electromagnetic (fast Alfv\'en) component. 

When the wave propagation angle with respect to the magnetic field is 89$^{\circ}$ rather than being strictly perpendicular to it (figure 12b), there is a greater transfer of energy late in the simulation from energetic ions to bulk ions. The transfer of energy to electrons is also somewhat greater in this case, while asymptotically the fluctuating field energy is found to be negligible. Stronger suppression of the fluctuations is to be expected in this case, since the presence of a finite parallel wavenumber $k_{\parallel}$ introduces the possibility of electron Landau damping. Indeed we find that when $\theta$ is reduced further the MCI becomes progressively weaker due to electron damping, thereby constraining the values of $k_{\parallel}$ that are compatible with the occurrence of ICE. 

Even in the case of oblique propagation, it is clear from figure 12 that the energy transfer resulting from the MCI is principally from suprathermal ions into bulk ions. In contrast, a purely collisional interaction between the ring-beam and bulk plasma used in these simulations would result in most of the ring-beam energy being transferred to electrons, since the ring-beam energy exceeds the critical value above which fast ions collide more frequently with electrons than with bulk ions \cite{Stix1975}. The former process is one of the essential features of proposed ``$\alpha$-channeling'' scenarios in which waves are used to transmit energy from fusion $\alpha$-particles directly into fuel ions, while simultaneously transporting the $\alpha$-particles radially, thereby facilitating the removal of helium ash from burning plasmas \cite{Fisch1994,Cook2010,Cook2017}. Figure 12 shows explicitly that a process akin to $\alpha$-channeling can occur spontaneously in regions of the plasma with fast ion velocity distributions that are not monotonic decreasing. The aymptotic change in bulk ion energy shown in this figure corresponds to an ion temperature increase of several tens of eV, which is a significant fraction of the initial temperature (100$\,$eV). Such temperature rises are in principle detectable, but unfortunately ion temperature measurements are available only for the core plasma in JET pulse \#79363, out to a major radius (3.79$\,$m) that lies several cm inside the midplane separatrix. These measurements do not show any significant temperature increase at the time of ICE detection. It is possible that bulk ion heating due to the MCI could be occurring closer to the plasma edge. The density profile shown in figure  the fast ion fraction ($10^{-3}$) in the {\tt EPOCH} simulations is unrealistically high. A lower fast ion fraction would reduce the MCI growth rate and the absolute amount of energy transferred to the bulk ions, but would not be expected to change significantly the partition of free energy between particle species and fields. The rapid growth resulting from the use of a high fast ion concentration makes it possible to access the nonlinear regime in a relatively short time, which in turn makes it possible to use very large numbers of particles, providing a high signal-to-noise ratio.

The modes excited in the simulations can be clearly identified by Fourier transforming the fluctuating fields in both space and time, thereby generating a dispersion plot. This is shown in figure 13 for the case of perpendicular propagation (the results for $\theta = 89^{\circ}$ are very similar). Specifically, this figure shows the energy in magnetic field ($B_z$) fluctuations, Fourier transformed in space over the simulation domain and in time over ten energetic ion cyclotron periods. The dispersion plot is dominated by a linear feature with phase velocity close $c_A$, i.e. a fast Alfv\'en wave. Within this linear feature, local maxima can be seen in the electrostatic field energy at frequencies close to the $^3$He cyclotron frequency (highlighted by a dashed line) and many of its harmonics. Lines of approximately constant frequency close to the cyclotron harmonics can be identified as ion Bernstein waves. The fluctuating magnetic energy actually peaks at frequencies in the lower hybrid range, which is not included in figure 13: this has been observed in earlier {\tt EPOCH} simulations with similar initial conditions \cite{Cook2010,Cook2011}.   

It was not possible to obtain wavenumber measurements of the ICE described in section 2; spectral information is available only in the frequency domain. For this reason it is useful to plot the fluctuating magnetic field energy as a function of frequency alone, and to compare this quantity with the corresponding results obtained in the absence of a ring-beam. Such a comparison, which indicates the extent to which the presence of the ring-beam boosts the wave energy above the noise level across a range of frequencies, is made in figure 14. The propagation angle was taken to be 90$^{\circ}$. In the case of the simulation with the ring-beam, the parallel drift was taken to be zero, but for this propagation angle the results are essentially independent of $v_{\parallel 0}$. The $^3$He cyclotron frequency is again highlighted by a dashed line. It is apparent from this figure that the presence of the ring-beam boosts the fluctuation power at $^3$He cyclotron harmonics, including the fundamental (again highlighted by a dashed line) by a factor of around two orders of magnitude. We note that the fluctuation energy associated with the fundamental peaks at a frequency slightly lower than $\Omega_{^3\rm{He}}$, which is consistent with the spectrum measured using the SHAD system in JET pulse \#79363 if the excitation region is the low field side plasma edge. The results obtained using this PIC simulation are thus broadly consistent with the ICE data, insofar as they show wave excitation at frequencies close to experimentally-measured values. As noted in section 3, ICE was only detected in pulse \#79363 after the disappearance of a large amplitude $n=2$ mode and a sharp drop in the rate of fast ion losses. It seems probable that the orbits of high energy $^3$He ions traversing the edge plasma region were sufficiently perturbed by the $n=2$ mode for the ions to be deconfined, and that it was only after the disappearance of this mode that the local distribution of confined $^3$He ions in the outer midplane edge could be approximated by equations (1) and (2).        

\subsection{Hybrid simulations}

Results very similar to those given by {\tt EPOCH} have been obtained in simulations performed using a one-dimensional hybrid code with identical initial conditions. The hybrid simulation boxes consisted of 8,192 cells, with 8,000 bulk ions per cell and 8,000 energetic ions per cell; the cell size was comparable to the bulk ion Larmor radius. This code can be run with or without electron inertial effects taken into account (the electron mass is always finite in {\tt EPOCH} simulations, of course, since this is a full PIC code). The scheme used in the case of finite electron mass is similar to that described in Ref. \cite{Amano2014}. It was found that electron inertia had no significant effects on the results of the simulations. We have also used this code to test the sensitivity of the MCI to the initial spread of velocities in the $^3$He fast ion distribution. Instead of equation (1), the fast ions were prescribed to have an initial distribution of the form
$$ f \sim \exp\left[-{(v_{\parallel}-v_{\parallel 0})^2\over v_{r\parallel}^2}\right] \exp\left[-{(v_{\perp}-v_{\perp 0})^2\over v_{r\perp}^2}\right], \eqno (3) $$ 
with $v_{r\parallel}$ and $v_{r\perp}$ set equal to specified fractions of $v_{\parallel 0}$ and $v_{\perp 0}$. Results from a simulation with massless electrons and $v_{r\parallel} = 0.1v_{\parallel 0}$, $v_{r\perp} = 0.1v_{\perp 0}$ are shown in figure 15. The upper frame of this figure shows the temporal evolution of the energy densities in the particle species and field components, while the lower frame shows the spectrum of fluctuations obtained by Fourier transforming the fields over the entire duration of the simulation (twenty $^3$He cyclotron periods). As in the {\tt EPOCH} simulation with a $\delta$-function initial fast ion distribution (figure 12), the energy components approach steady-state values in the nonlinear phase of the MCI. The time taken for this to happen is somewhat longer in the case of the hybrid simulation, but the asymptotic energy densities of the fast ions and bulk ions in figure 15a are remarkably close to those in the upper frame of figure 12, indicating that the nonlinear evolution of the instability is insensitive to both electron dynamics and the initial velocity spread of the fast ions. The longer time required to reach saturation in the hybrid case is due entirely to the choice of fast ion distribution rather than the use of a fluid model for the electrons: in a hybrid simulation with an initial fast ion distribution given by equation (1), the energy components evolved on timescales that were essentially identical to those in the equivalent {\tt EPOCH} simulation (figure 11a).   

The finite bandwidths of the modes associated with cyclotron harmonics are again apparent in the spectrum, figure 15b. Like figure 14, this plot shows the spectrum of fluctuations in $B_z$, and thus makes explicit the electromagnetic character of the MCI. The mode at $\omega \simeq \Omega_{^3{\rm He}}$ in figure 15b is about an order of magnitude more intense than it is in a spectrum generated by using only data from the linear phase of the instability. This illustrates the importance for ICE interpretation of modelling the nonlinear stage of the MCI, a point which has been noted in previous studies carried out using the hybrid approach \cite{Carbajal2014}. As in the case of the particle-in-cell simulations, the excitation frequencies shown in figure 15b are broadly consistent with the ICE observations reported in section 3.     

\section{Conclusions and discussion}

\noindent We have presented measurements of ion cyclotron emission (ICE) from JET plasmas heated by a combination of ICRH and neutral beams; the measurements were obtained using a detection system in the ICRH antenna transmission lines. Typically the ICE was detected in plasmas with hydrogen as the dominant ion species and $^3$He as a minority species, the latter being accelerated by ICRF waves via the second harmonic cyclotron resonance. In most cases the detected ICE frequencies are close to the $^3$He cyclotron frequency at the outer midplane plasma edge, and in several pulses ICE detection was triggered during the ramp-up in ICRH power, strongly suggesting that the emission was driven by ICRF-accelerated $^3$He ions. However in one pulse (\#79363) excitation was delayed until a reduction in the neutral beam power and the suppression of a large amplitude $n=2$ mode whose presence was correlated with strong fast ion losses. In two other pulses the detected ICE frequency (around 15$\,$MHz) was intermediate between the edge plasma $^3$He cyclotron frequency (17.4$\,$MHz) and deuterium/$^4$He cyclotron frequency (13.1$\,$MHz), although it was somewhat closer to the latter. In all cases the measured spectra contain fine structure that appears to be intrinsic to the plasma rather than being an instrumental effect.   

PIC and hybrid simulations, performed using field, bulk plasma and fast particle parameters corresponding to the low field side plasma edge at the time of excitation of ICE in pulse \#79363, show wave excitation via the magnetoacoustic cyclotron instability at many $^3$He cyclotron harmonics, including the fundamental, which we identify with the measured emission. In the PIC simulations the wave power is typically two orders of magnitude above the noise level, the latter being quantified by performing an identical simulation without the ring-beam. The simulations also show a significant fraction of the ring-beam energy being channeled into bulk ions, with a much smaller energy transfer into electrons and fluctuating fields. The hybrid simulations demonstrate that the instability is robust with respect to the initial spread of fast ion velocities. In view of the gradients in background plasma density and temperature close to the plasma edge shown in figure 11, it is important to consider the sensitivity of the simulation results to these parameters. As noted previously, the results are insensitive to electron dynamics, and in fact they are also insensitive to the background ion temperature, $T_i$: plots of the changes in energy density of the particle species and fields in PIC simulations with initial $T_i = 50\,$eV, 100$\,$eV and 200$\,$eV are very similar. The results are somewhat more sensitive to the bulk ion density, since the Alfv\'en speed $c_A$ and hence the ratio of fast particle speed $v_f$ to $c_A$ depend on this parameter. When $v_f/c_A$ is reduced, the predicted growth rate of the MCI generally goes down, essentially because the fast particle terms in the dielectric tensor elements contain Bessel function factors whose arguments are proportional to this ratio \cite{Belikov1976}. However, it is important to note that strong instability does not require the fast ions to be super-Alfv\'enic (unlike, for example, the excitation of toroidal Alfv\'en eigenmodes (TAEs) via the primary resonance, $\omega = k_{\parallel}c_A$). This is demonstrated for example by hybrid simulations of ICE driven by sub-Alfv\'enic beams in LHD \cite{Reman2016}, and more recent unpublished simulations using LHD parameters show that ICE can be excited in this device for values of $v_f/c_A$ as low as 0.51.    

These results underline the potential importance of ICE measurements as a method of studying confined fast particles that are strongly suprathermal but have insufficient energies or are not present in sufficient numbers to excite detectable levels of $\gamma$-ray emission or other collective instabilities. If, however, the fast ion energies in the PIC and hybrid simulations had been 10$\,$MeV rather than 2$\,$MeV, as suggested by the {\tt PION} results (figure 9), we would have expected the MCI to be even more strongly driven. The detection of ICE cannot therefore be used on its own to determine the maximum fast ion energy.         

There are significant differences between the measured ICE spectrum in \#79363 (figure 5b) and the simulated spectra (figures 14 and 15b). First of all it should be recalled that the measured frequencies are strongly filtered by the response function of the SHAD system shown in figure 1: thus, in general, the frequency of highest signal intensity in the measured spectrum does not coincide with the peak emission frequency in the plasma. As noted previously, the spectrum shown in figure 5b peaks at a frequency (19.25$\,$MHz) that is slightly lower than the $^3$He cyclotron frequency at the outer midplane edge (20.7$\,$MHz). However, as shown in figures 14 and 15b, the instability at the fundamental observed in the PIC and hybrid simulations has a finite bandwidth, which extends down to the measured range. Given that $^3$He ions accelerated at the second harmonic resonance have a large parallel velocity in the outer midplane, we might expect the instability to be Doppler shifted from the fundamental, since the fundamental resonance condition for a uniform plasma is given by 
$$ \omega - k_{\parallel}v_{\parallel} - \Omega_{^3\rm{He}} = 0. \eqno (4) $$ 
However, as noted in the previous section, our PIC simulations show strong instability occurring only for wave propagation angles close to 90$^{\circ}$, and for the simulations with $\theta = 89^{\circ}$ and $v_{\parallel 0} =  5\times  10^6\,$ms$^{-1}$ the Doppler shift indicated by equation (4) is about 125$\,$kHz, which is less than 1\% of the emission frequency. The finite parallel velocities of the fast ions are thus unlikely to produce significant deviations of the ICE frequency from the fundamental resonance.   

A rather larger frequency shift might be expected to arise from modifications to the resonance condition resulting from magnetic field gradients and curvature. In a toroidal plasma it can be shown that equation (4) has the modified form \cite{Gorelenkov1995,Fulop1997}         
$$ \omega - k_{\parallel}v_{\parallel} - \Omega_{^3\rm{He}} - \omega_D = 0. \eqno (5) $$ 
where $\omega_D $ is the fast ion grad-$B$/curvature drift frequency. In the large aspect ratio limit this is equal to $-mv_{\nabla B}/r$ where $m$ is poloidal mode number, $r$ is minor radius and $v_{\nabla B} = (v_{\parallel}^2+v_{\perp}^2/2)/\Omega_{^3\rm{He}}R$, $R$ being major radius. Evaluating $\omega_D$ for 2$\,$MeV $^3$He ions in the outer edge plasma of JET pulse \#79363, we obtain $\vert\omega_D\vert/2\pi \simeq 40m\,$kHz. As noted previously, it was not possible to determine mode numbers from the SHAD measurements in JET. However toroidal mode numbers $n$ of ICE have been obtained in JT-60U \cite{Ichimura2008} and, more recently, in the MAST spherical tokamak, where compressional Alfv\'en eigenmodes (CAEs) with frequencies in the ion cyclotron range (i.e. ICE eigenmodes) were found to have $n \simeq 8-12$ \cite{Sharapov2014}. In the case of fundamental $^3$He ICE in JT-60U ($R \simeq 3.4\,$m), Ichimura and co-workers recorded toroidal wavenumbers $k_{\varphi} = n/R \simeq 3\,$m$^{-1}$, i.e. $n \simeq 10$. In this case the $^3$He ions were fusion products rather than ICRF-accelerated fast ions, but, given the similarities in geometry and scale between JT-60U and JET, it  is reasonable to expect fundamental $^3$He ICE in the two devices to have similar mode numbers. Moreover, given that we expect ICE to have parallel wavenumbers $k_{\parallel}$ close to zero, for the reasons discussed in section 4, it follows from the large aspect ratio expression $k_{\parallel} \simeq (n-m/q)/R$ that $m \simeq nq$ for these modes. Uncertainty in the precise location of the ICE source region in JET pulse \#79363, combined with the steep profile of $q$ in the edge plasma, makes it difficult to estimate the local value of this quantity, but one would expect $q \gg 1$ in this region and therefore that the ICE is characterised by $m \gg n$, i.e. $m \gg 10$ if the mode numbers of fundamental $^3$He ICE in JT-60U and JET are comparable. The expression $\vert\omega_D\vert/2\pi \simeq 40m\,$kHz could thus yield frequency shifts of the order of 1$\,$MHz, significantly offsetting ICE frequencies from those expected in the case of a uniform equilibrium plasma.   

The possibility of such frequency shifts and the absence of mode number measurements also complicate the interpretation of the ICE detected in pulse \#79365. However, as discussed at the end of section 3, a possible explanation of the emission in this case is that it was driven by $^4$He ions accelerated by ICRF waves via the third harmonic resonance. 

Another significant difference between the measured and modelled ICE spectra is the presence of fine structure in the former and its absence from the latter. It is possible that the peaks in the measured spectra could be due to the excitation of separate CAEs, which cannot be captured by 1D3V PIC or hybrid simulations. The frequency spacing of the peaks (of the order of 10$\,$kHz) is very small compared to the emission frequency ($\sim 20\,$MHz). We note that CAEs with a similar absolute frequency spacing have been detected in MAST pulses with low toroidal field (see figure 5 in Ref. \cite{Sharapov2014}). However an unambiguous identification of the peaks in figures 5 and 6 as individual CAEs is again hampered by the lack of mode number measurements.    

The SHAD measurements demonstrate that ICE can be detected at relatively low cost using hardware that lies entirely outside the vacuum vessel. An alternative method of detection, which is expected to be available in both forthcoming JET campaigns and in ITER, is microwave reflectometry \cite{McClements2015}: the efficacy of this technique for studies of CAEs in the ion cyclotron range has been demonstrated in the NSTX spherical tokamak \cite{Crocker2011}. The measurements and modelling presented above reinforce the important point that ICE is a collective instability which is often driven by highly suprathermal (but not necessarily super-Alfv\'enic) fast ions in fusion plasmas. It provides an experimentally-detectable signature of fusion $\alpha$-particle effects in deuterium-tritium plasmas that is complementary to other such signatures, for example $\alpha$-particle-driven TAE excitation, and for this reason should if possible be studied in future deuterium-tritium campaigns.  

\section*{Acknowledgments}

\noindent This work has been carried out within the framework of the EUROfusion Consortium and has received funding from the Euratom research and training programme 2014-2018 under grant agreement No 633053 and from the RCUK Energy Programme [grant number EP/P012450/1]. The views and opinions expressed herein do not necessarily reflect those of the European Commission. AB was supported by the Embassy of France in the United Kingdom. We thank Ernesto Lerche for helpful information on the determination of $^3$He concentration in the pulses discussed in the paper.

\section*{ORCID iDs}

\noindent B. Chapman  https://orcid.org/0000-0001-9879-2285

\noindent S. C. Chapman https://orcid.org/0000-0003-0053-1584

\noindent K.G. McClements https://orcid.org/0000-0002-5162-509X

\newpage

\begin{figure}[ht]
\includegraphics[clip=true, trim = 0.0cm -2.0cm 0.0cm -6.0cm, width = 16.0cm]{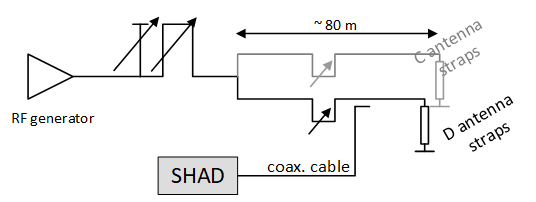}
\caption{\label{fig:SHAD_diagram} Diagram showing the integration of SHAD in the JET A2 C and D antenna system. C and D antennas can be fed independently, or in the so-called conjugate T configuration where C (shown in grey) and D antenna straps are fed in parallel.} 
\end{figure}

\begin{figure}[ht]
\includegraphics[clip=true, trim = 0.5cm 3.5cm 0.0cm 3.5cm, width = 16.0cm]{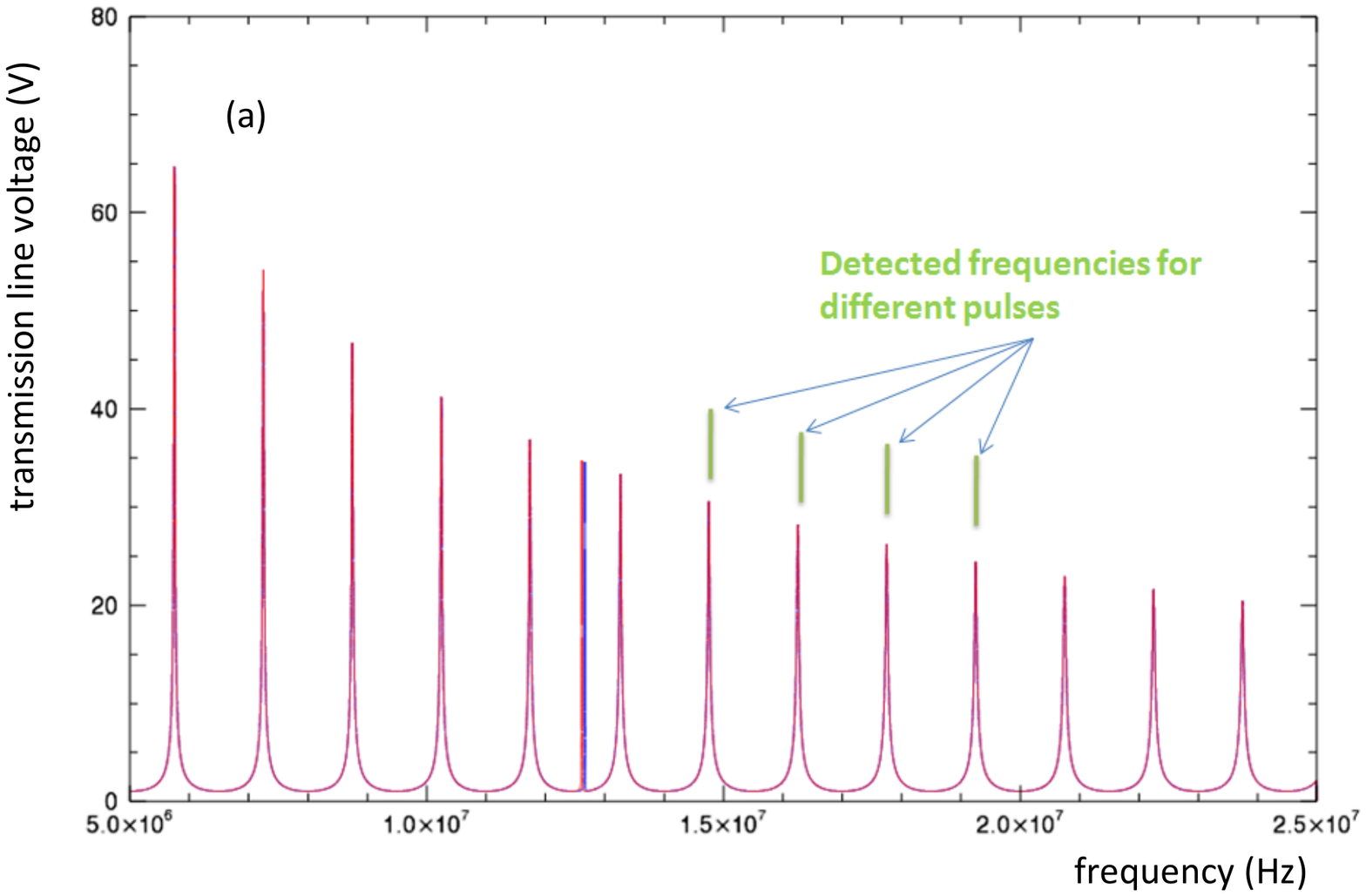}
\includegraphics[clip=true, trim = 3.0cm 3.0cm 0.0cm 3.5cm, width = 16.0cm]{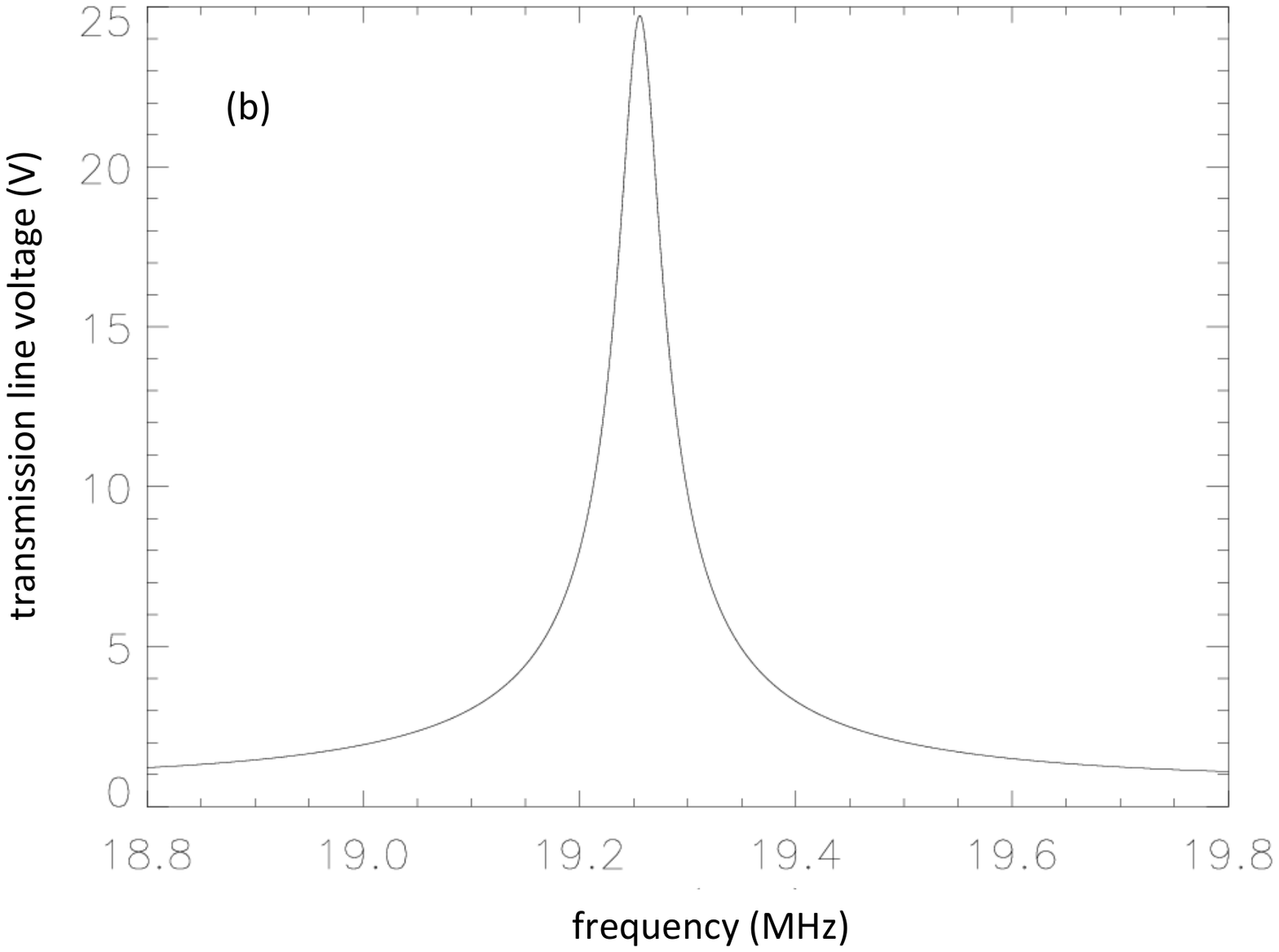}
\caption{\label{fig:cavity_response} (a) Calculated  voltage in the transmission line for a 1$\,$V signal entering antenna D straps (C and D fed independently) with the matching elements tuned for operation at 51.53$\,$MHz. The green bars show the frequencies at which ion cyclotron emission was recorded in a series of pulses with the same antenna system configuration. (b) Zoom on the peak at 19.2 MHz. The width of the peak (which is related to the quality factor of the RF cavity), $\sim$80$\,$kHz in this case, depends essentially on the losses at the antenna straps.} 
\end{figure}

\begin{figure}[ht]
\includegraphics[clip=true, trim = 2.5cm 0.2cm -2.4cm 2.5cm, width = 18.0cm]{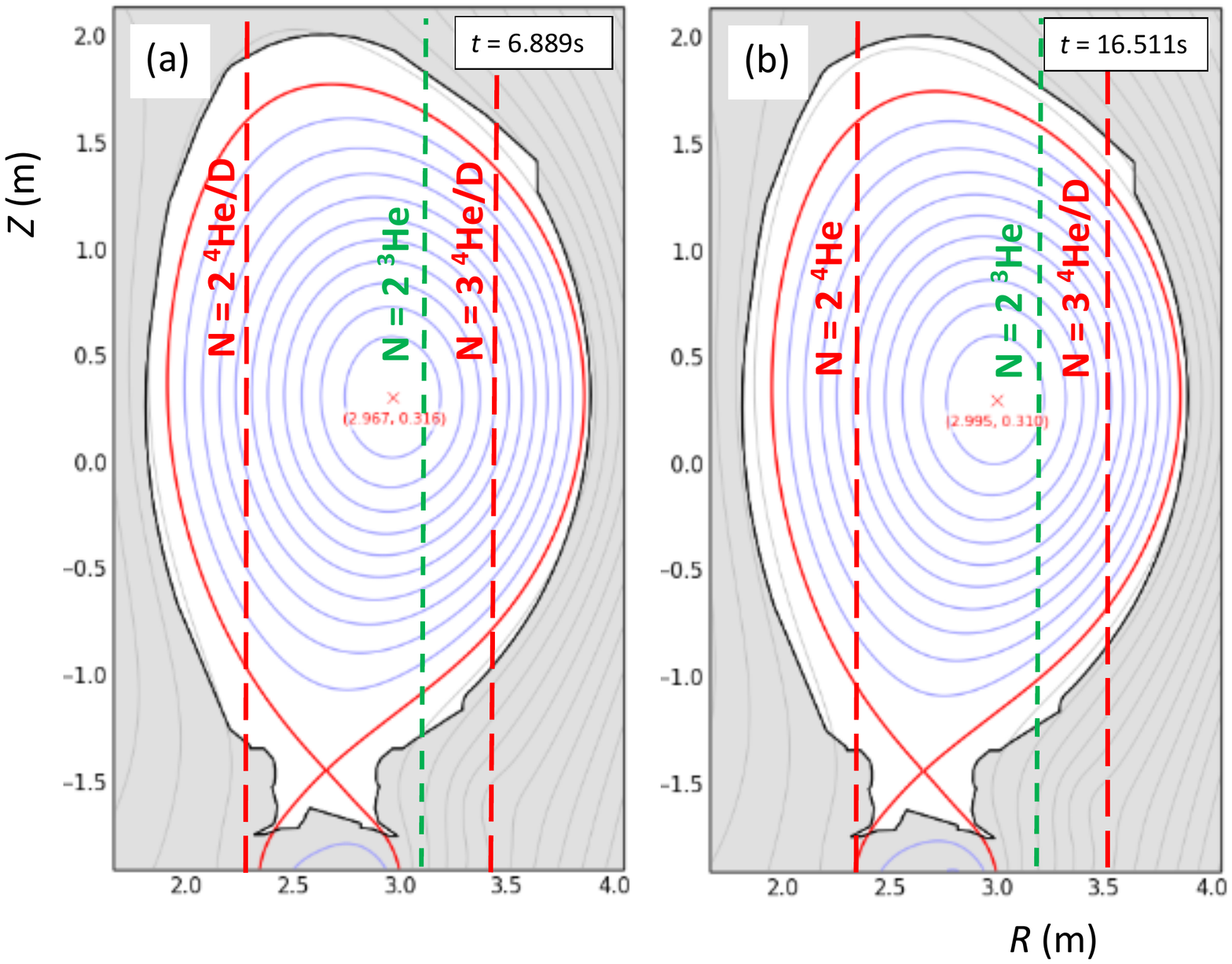}
\caption{\label{fig:spectrum2} Radial loci corresponding to second harmonic $^3$He (green dashed lines) and second/third harmonic $^4$He/D ICRF resonances (red dashed lines) in JET pulses (a) \#79363 and (b) \#79365. In both plots the equilibrium flux surfaces (blue curves) and separatrix (red solid curves) correspond to the times of ICE detection in these pulses.}
\end{figure}

\begin{figure}[ht]
\includegraphics[clip=true, trim = -1.0cm 4.5cm 1.0cm 4.5cm, width = 15.0cm]{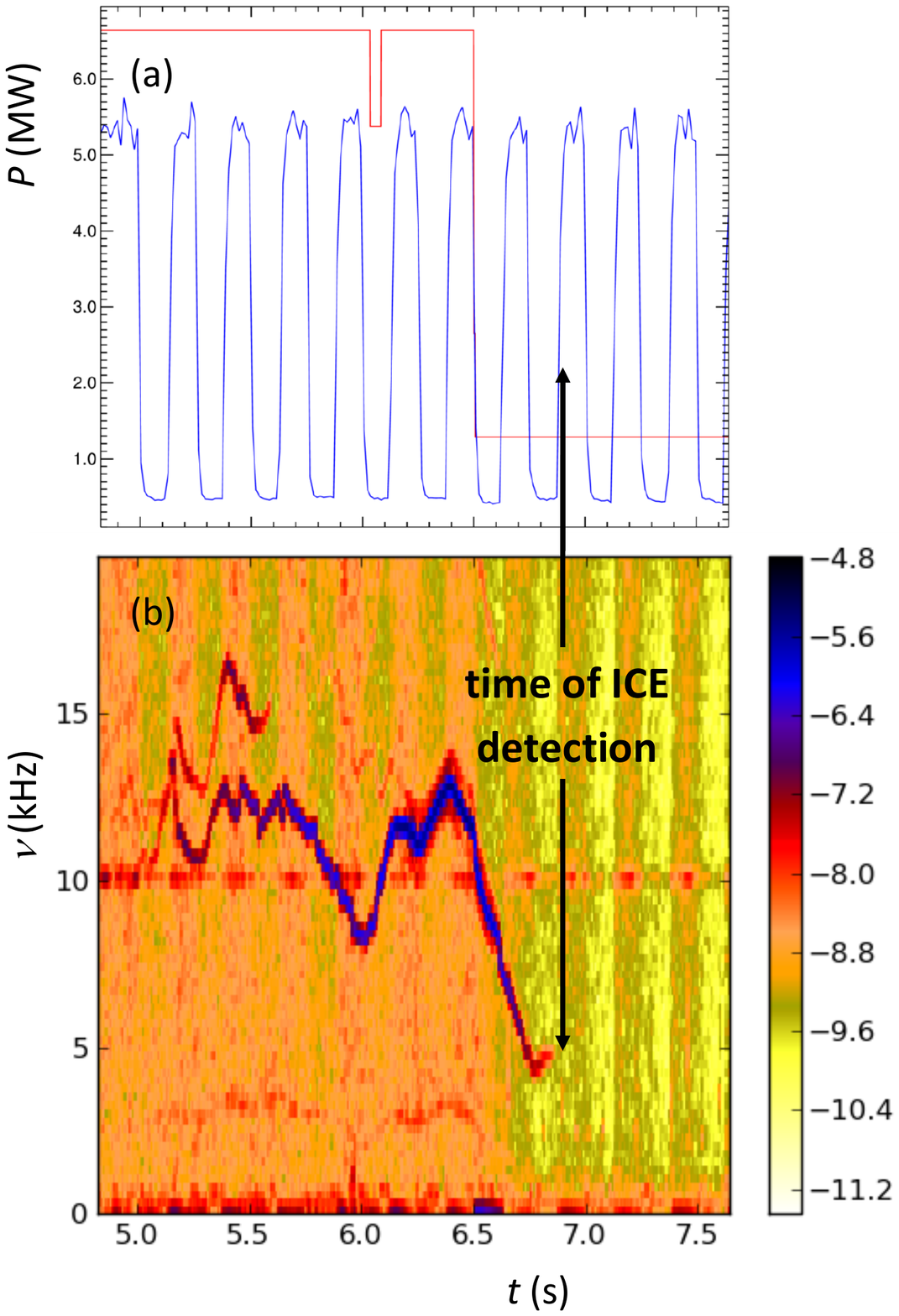}
\caption{\label{fig:MHD} (a) Coupled radio frequency power (blue) and neutral beam power (red), and (b) spectrogram of magnetic fluctuations shortly before and during the period of ICE in JET pulse \#79363. The colour scale in (b) provides a logarithmic measure of the measured fluctuating field magnitude. The mode shown in this plot had dominant toroidal mode number $n=2$ and was located inside a flux surface with safety factor $q=2$.} 
\end{figure}

\begin{figure}[ht]
\includegraphics[clip=true, trim = 2.0cm 2.0cm -2.0cm 5.0cm, width = 19.0cm]{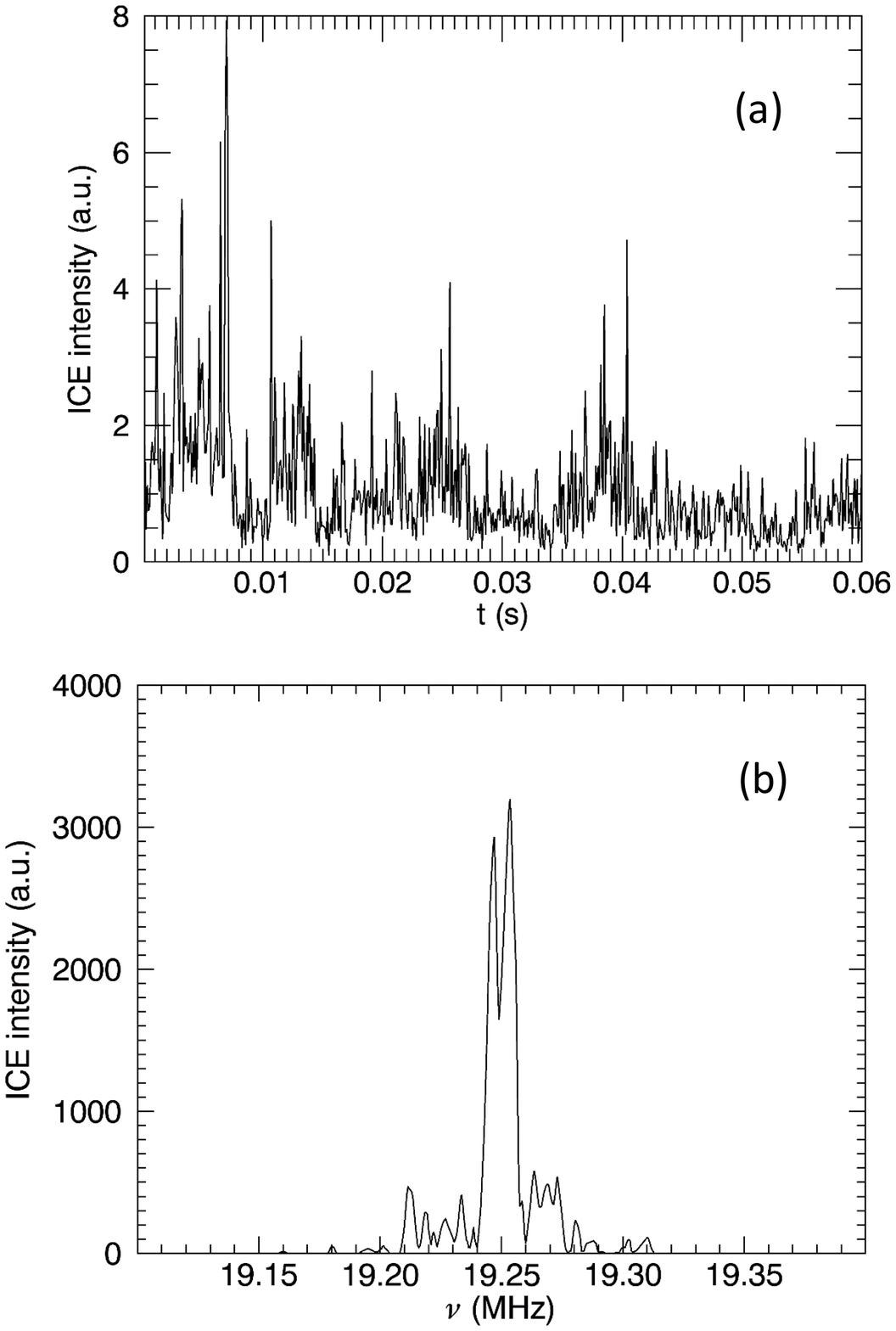}
\caption{\label{fig:spectrum1} (a) Temporal evolution of ICE intensity recorded using the SHAD system following the triggering of ICE detection in JET pulse \#79363. Time is measured from the triggering of ICE detection at 6.9$\,$s. (b) ICE spectrum at $t \simeq 6.94\,$s.} 
\end{figure}

\begin{figure}[ht]
\includegraphics[clip=true, trim = 2.0cm 2.5cm -2.3cm 4.0cm, width = 19.0cm]{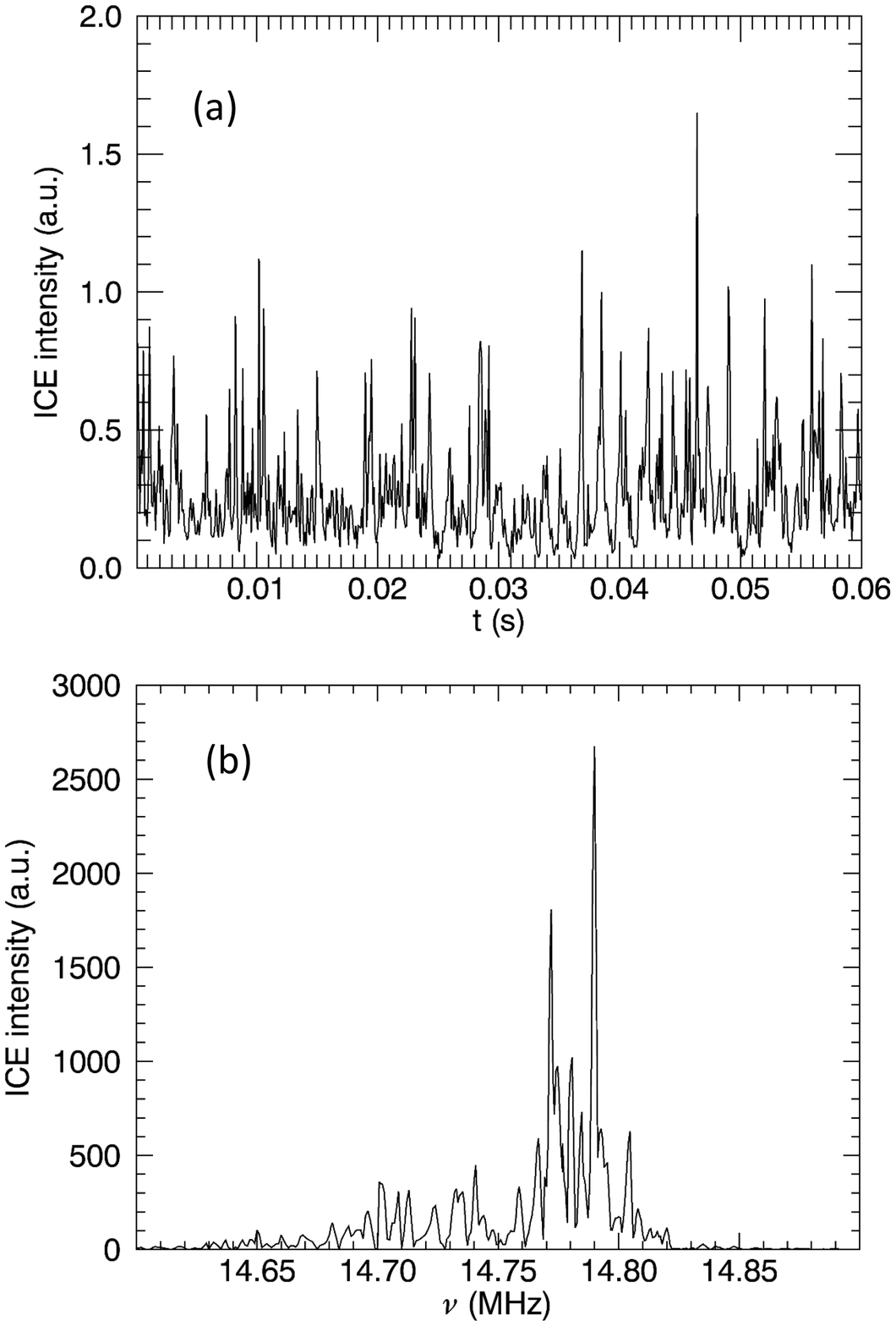}
\caption{\label{fig:spectrum2} (a) Temporal evolution of ICE intensity recorded using the SHAD system following the triggering of ICE detection in JET pulse \#79365. Time is measured from the triggering of ICE detection at 16.5$\,$s. (b) ICE spectrum at $t \simeq 16.51\,$s.} 
\end{figure}

\begin{figure}[ht]
\includegraphics[clip=true, trim = 2.0cm 1.2cm -2.0cm 3.3cm, width = 19.0cm]{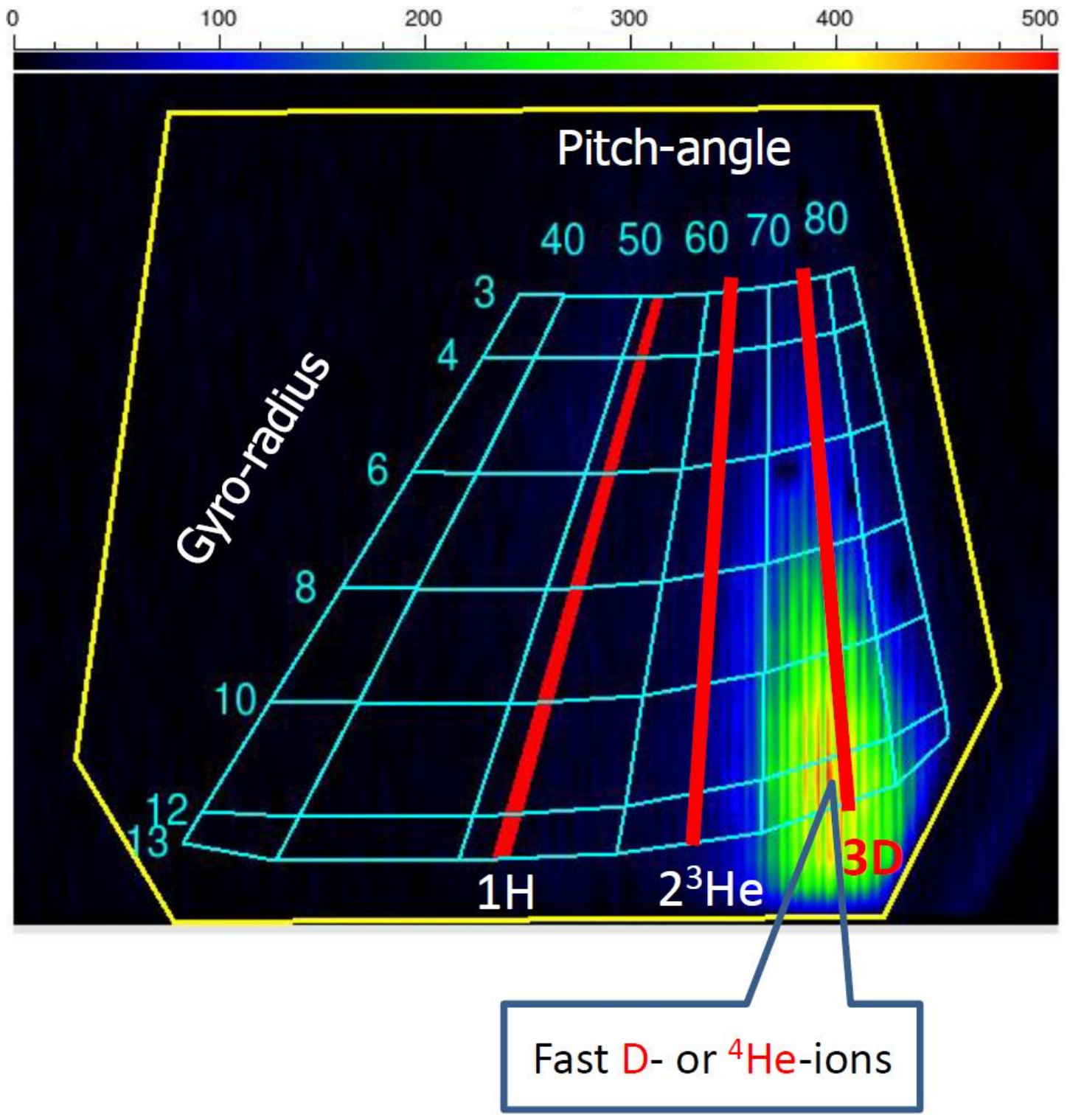}
\caption{\label{fig:spectrum2} Footprint of fast particle losses recorded using the Scintillator Probe plate in JET pulse \#79365 approximately 4$\,$s after the initial detection of ICE in this pulse. The red lines indicate the pitch angles of hydrogen, $^3$He ions and $^4$He/D ions accelerated by the applied ICRF waves via the fundamental, second harmonic and third harmonic resonances respectively.}
\end{figure}

\begin{figure}[ht]
\includegraphics[clip=true, trim = 0.0cm 8.0cm -4.0cm 0.0cm, width = 21.0cm]{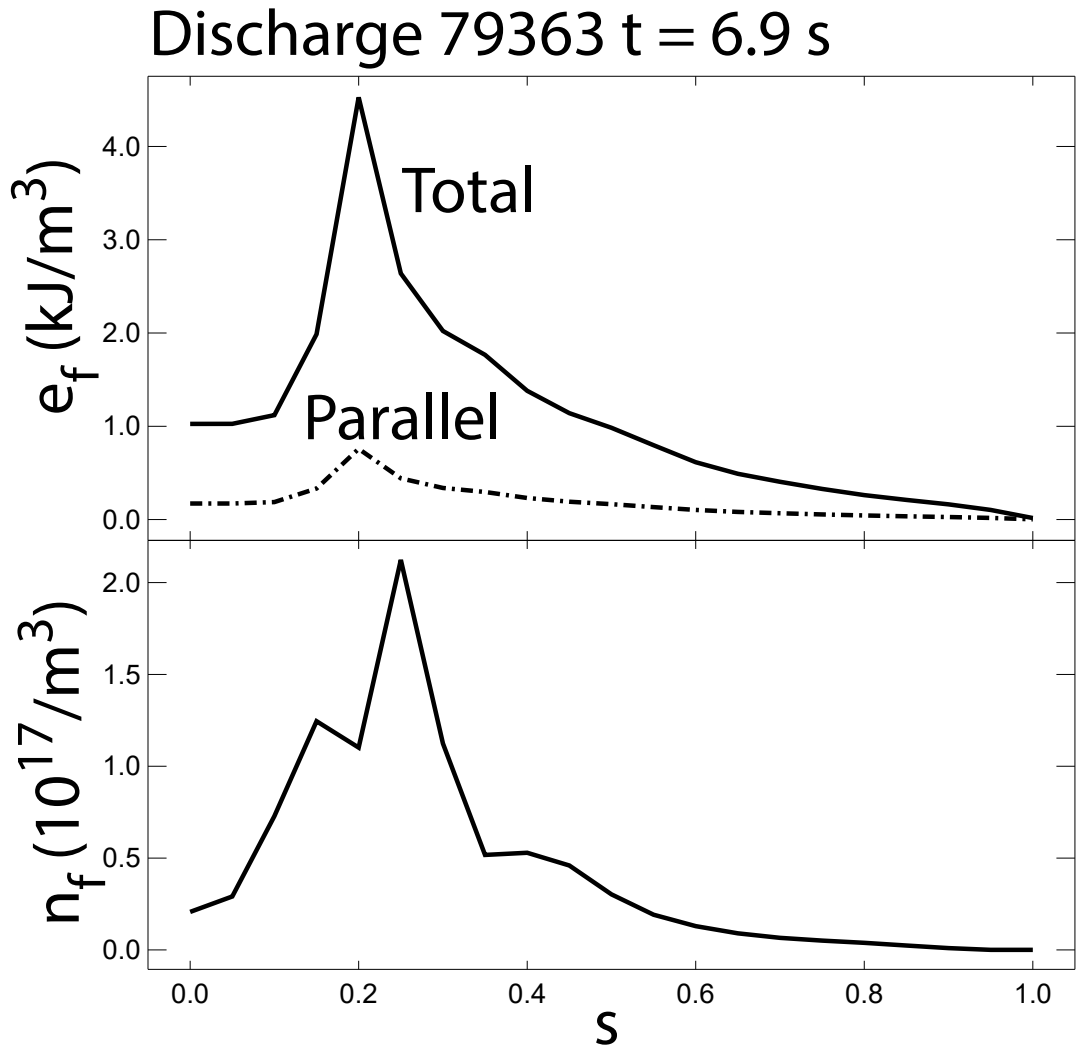}
\caption{\label{fig:spectrum2} Radial profiles of total (solid curve) and parallel (dashed curve) $^3$He fast ion energy density (top), and fast ion number density (bottom), computed using {\tt PION} for JET pulse \#79363, $t = 6.9\,$s. The total $^3$He density was assumed to be 7.1\% of the electron density, $n_e$.}
\end{figure}

\begin{figure}[ht]
\includegraphics[clip=true, trim = 2.5cm 10.0cm -0.5cm 3.0cm, width = 19.0cm]{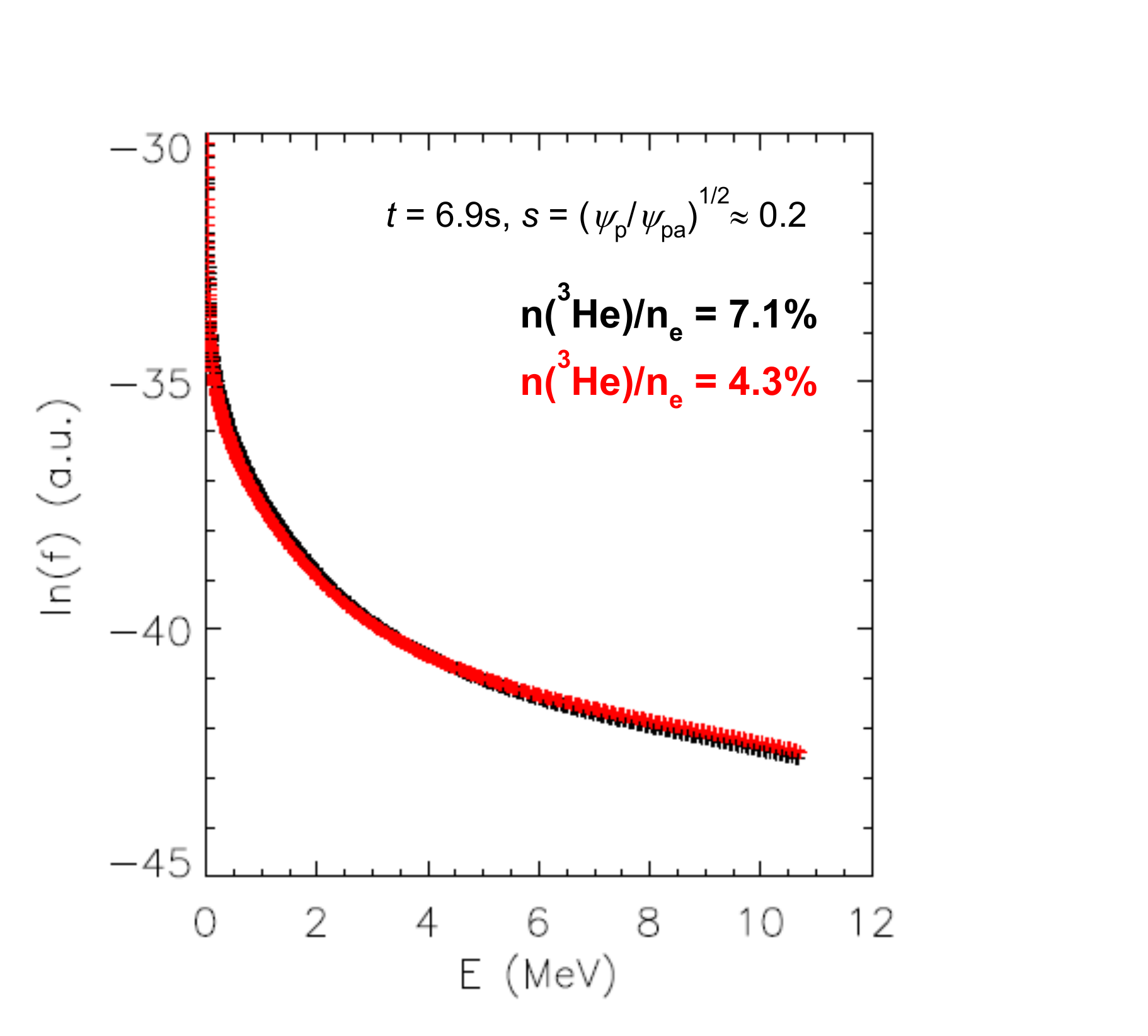}
\caption{\label{fig:spectrum2} $^3$He energy distributions for assumed total $^3$He concentrations of 4.3\% (red curve) and 7.1\% (black curve) computed using {\tt PION} for JET pulse \#79363, $t = 6.9\,$s, at normalised minor radius $s \simeq 0.2$.}
\end{figure}

\begin{figure}[ht]
\includegraphics[clip=true, trim = 2.5cm 0.2cm 0.0cm 2.5cm, width = 18.0cm]{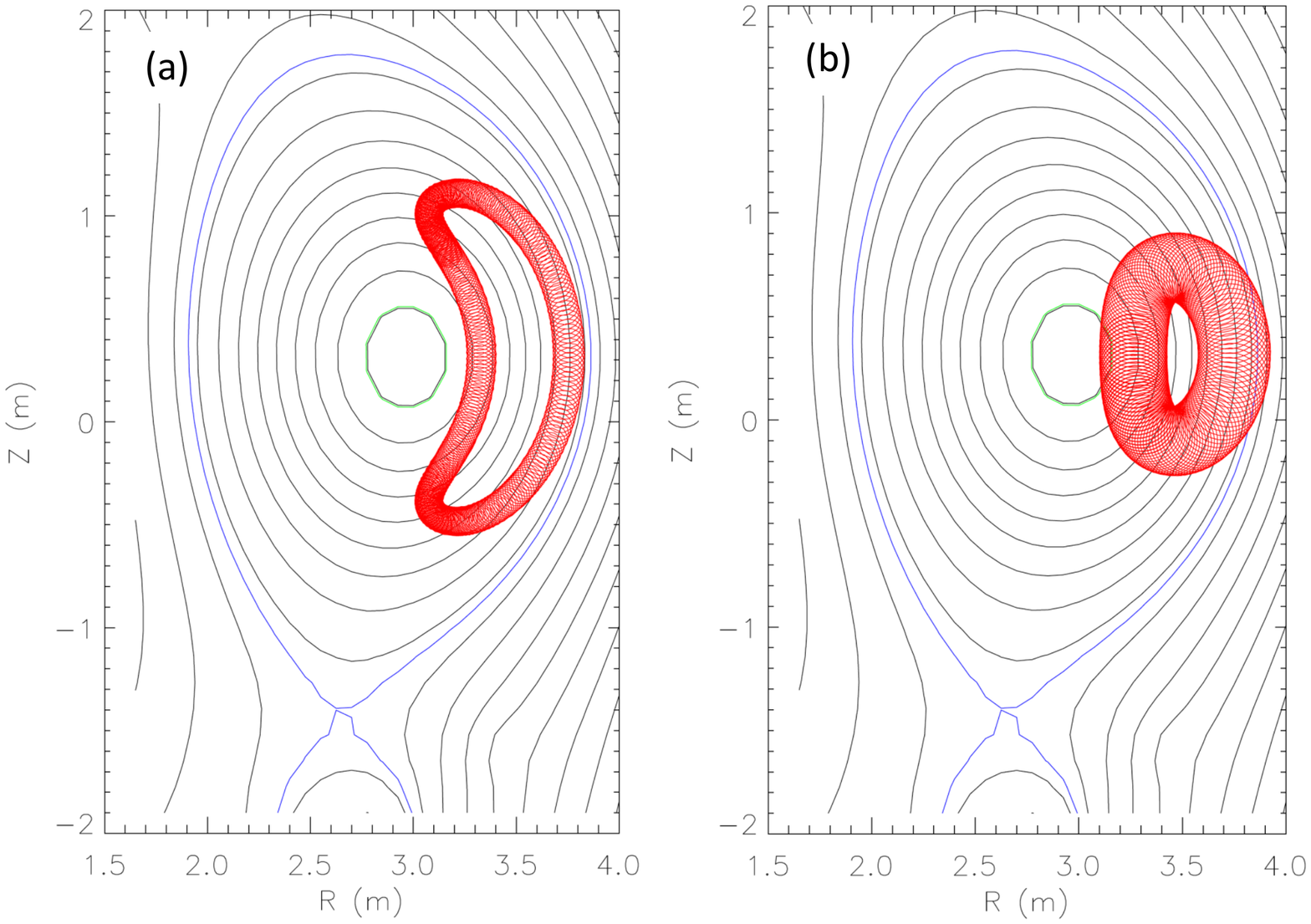}
\caption{\label{fig:spectrum2} Orbits projected onto the poloidal plane of $^3$He ions initially at $R =3.1\,{\rm m} \simeq  R_{\rm res}$ and with purely vertical velocities in the reconstructed equilibrium magnetic field of JET pulse \#79363, $t = 6.9\,$s (red curves). The particles have energies and initial mormalised minor radii of (a) $E = 2\,$MeV, $s = 0.66$, and (b) $E = 10\,$MeV, $s = 0.16$. The black contours are flux surfaces, the blue curves indicate the confined plasma boundary, and the green curves show the estimated location of the $q=3/2$ surface.}
\end{figure}

\begin{figure}[ht]
\includegraphics[clip=true, trim =1.5cm 2.2cm -2.0cm 1.5cm, width = 18.0cm]{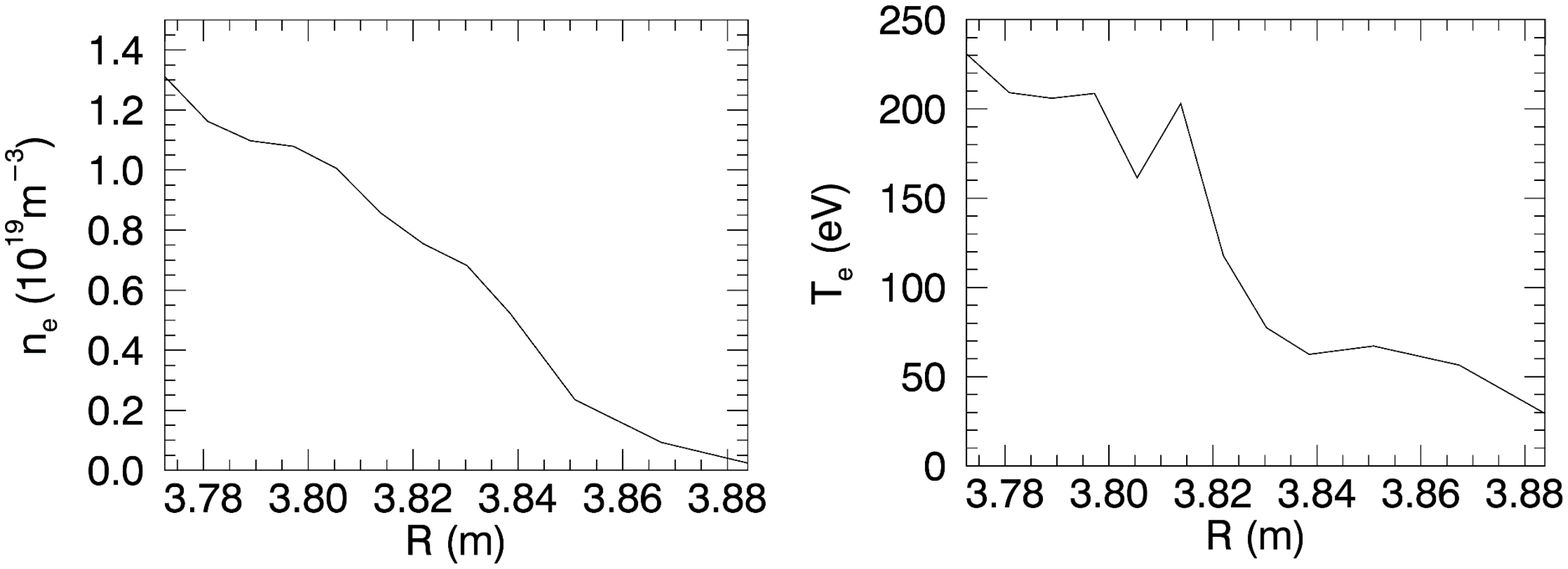}
\caption{\label{fig:HRTS} Thomson scattering measurements of electron density (left) and electron temperature (right) profiles close to the low field side plasma edge in JET pulse \#79363, $t \simeq 6.9\,$. The measurements have been averaged over a 400$\,$ms time interval to reduce noise.}
\end{figure}

\begin{figure}[ht]
\includegraphics[clip=true, trim = 0.5cm 4.5cm 1.5cm 3.5cm, width = 16.0cm]{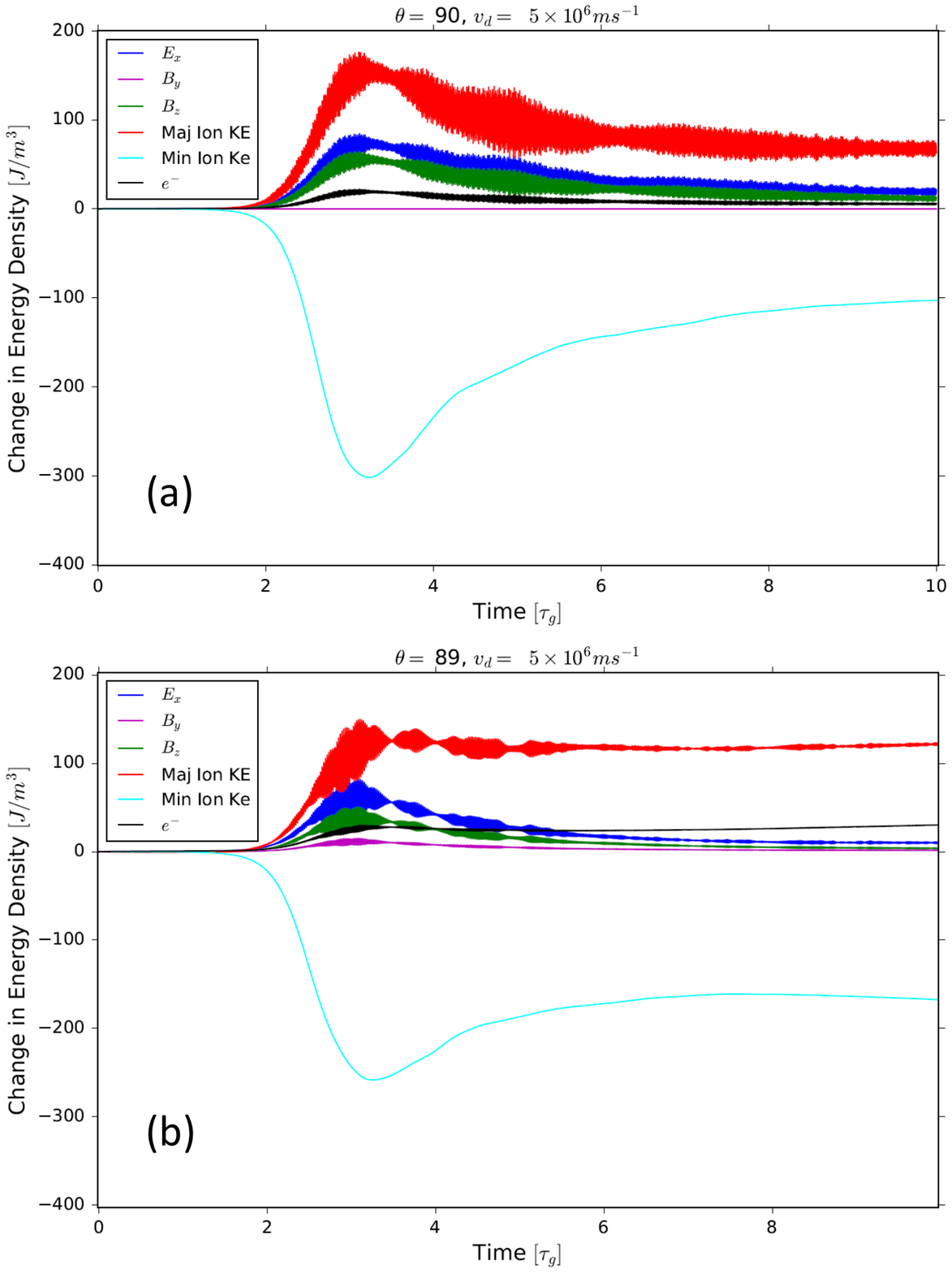}
\caption{\label{fig:energy} Changes in the energy densities of particle species (bulk protons in red, energetic $^3$He ions in turquoise, electrons in black) and field components ($E_x$ in blue, $B_y$ in magenta, $B_z$ in green) during {\tt EPOCH} simulations with wavevector inclined at an angle of (a) 90$^{\circ}$ with respect to $\bm{B}_0$ and (b) 89$^{\circ}$ with respect to $\bm{B}_0$. Time is in units of the $^3$He cyclotron period.} 
\end{figure} 

\begin{figure}[ht]
\includegraphics[clip=true, trim = 0.5cm 0.5cm 1.5cm 2.0cm, width = 16.0cm]{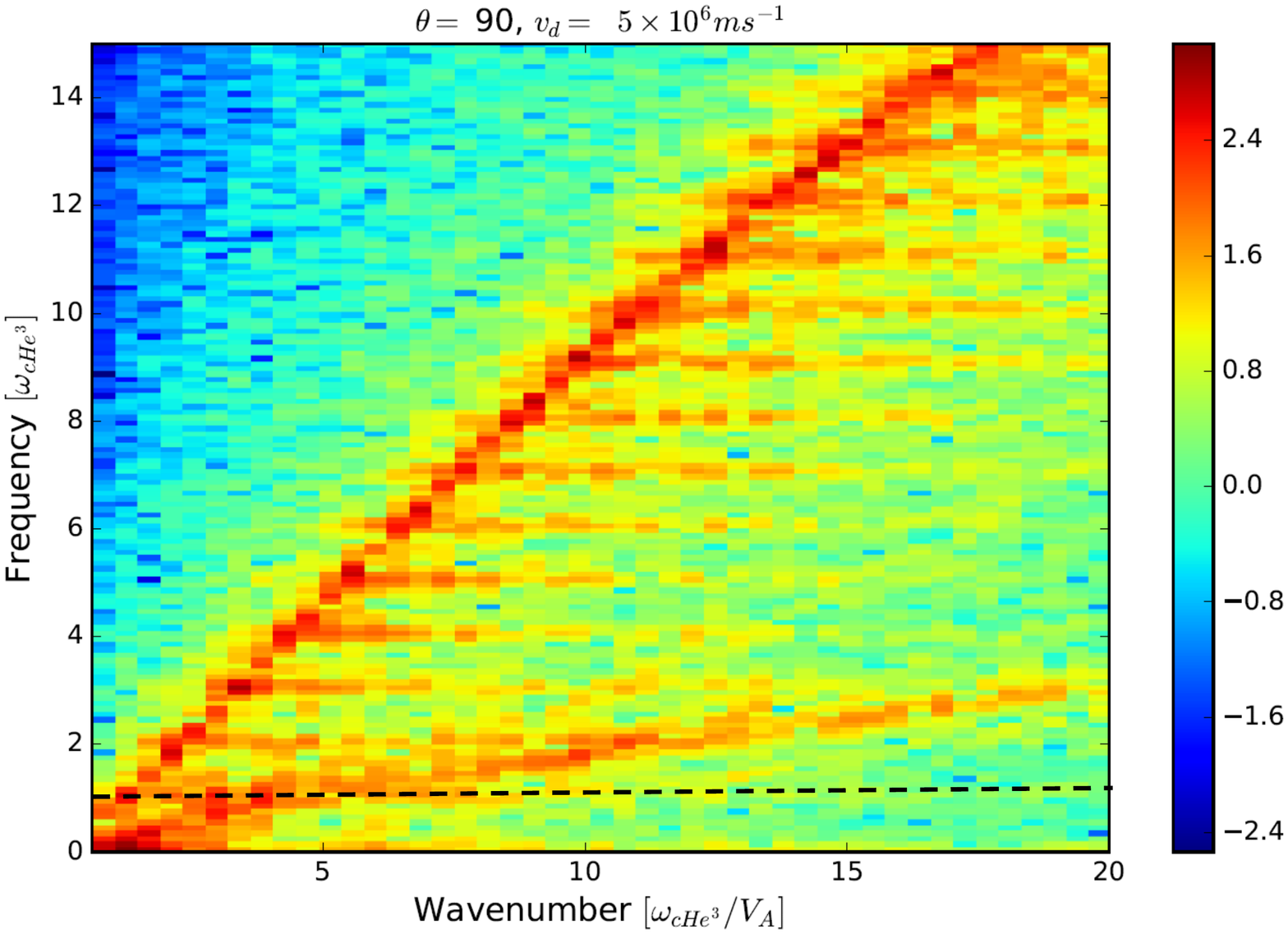}
\caption{\label{fig:dispersion} Power of magnetic field ($B_z$) fluctuations in wavenumber/frequency space (base 10 logarithmic scale, arbitrary units) obtained by Fourier transforming {\tt EPOCH} simulation results over the entire spatial domain and over ten  $^3$He cyclotron periods in the time domain. Frequency is in units of the $^3$He cyclotron frequency $\Omega_{^3\rm{He}}$, highlighted by a horizontal dashed line, and wavenumber is in units of $\Omega_{^3\rm{He}}/c_A$ where $c_A$ is the Alfv\'en speed.} 
\end{figure} 

\begin{figure}[ht]
\includegraphics[clip=true, trim = 0.5cm 0.5cm 0.5cm 0.5cm, width = 16.0cm]{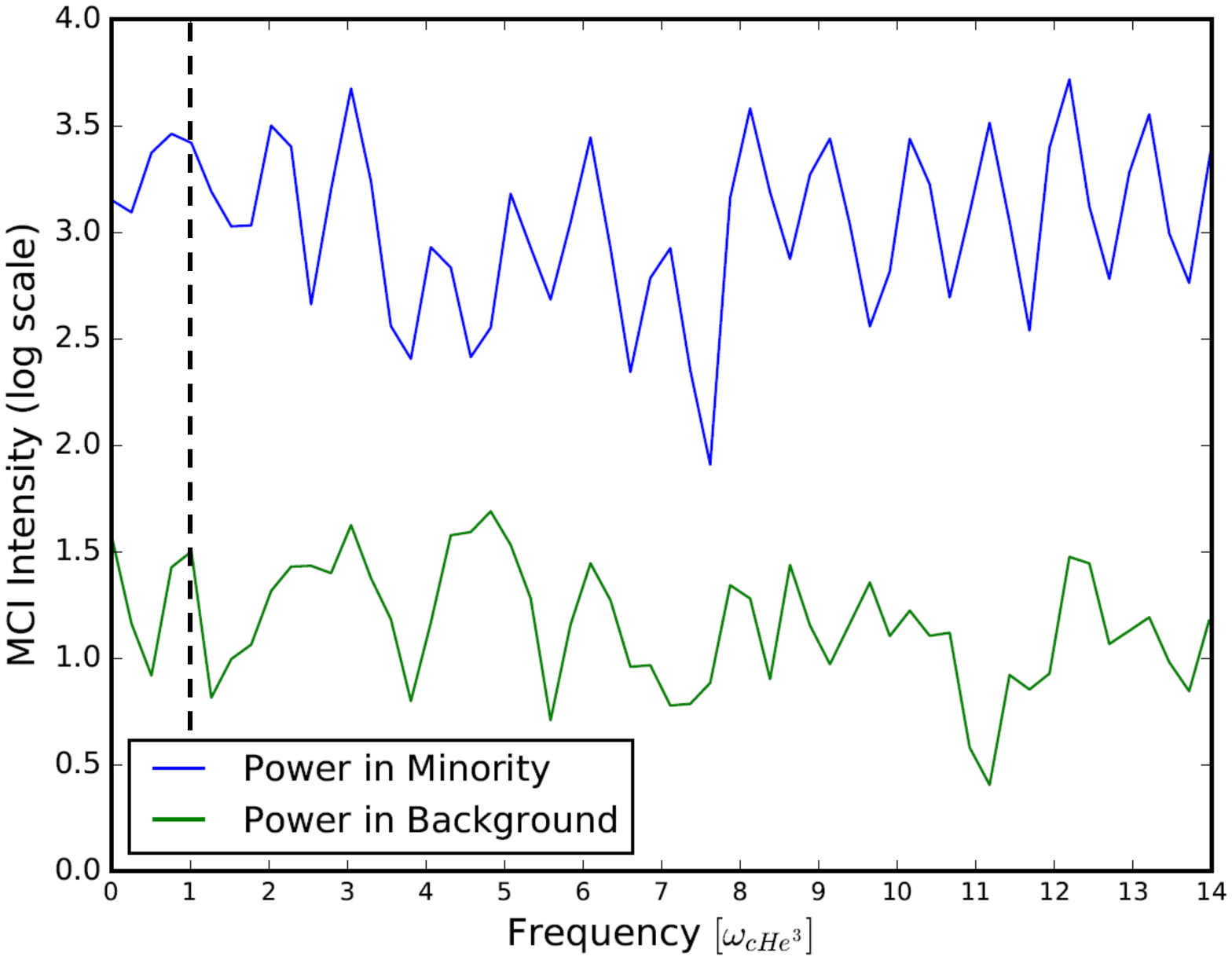}
\caption{\label{fig:power} Power of magnetic field ($B_z$) fluctuations in frequency space (arbitrary units) obtained by Fourier transforming field data over four $^3$He cyclotron periods in {\tt EPOCH} simulations with (upper curve) and without (lower curve) a $^3$He ring-beam. Frequency is in units of $\Omega_{^3\rm{He}}$: the fundamental is highlighted by a dashed line.} 
\end{figure} 

\begin{figure}[ht]
\includegraphics[clip=true, trim = 0.5cm 3.5cm 0.5cm 3.0cm, width = 16.0cm]{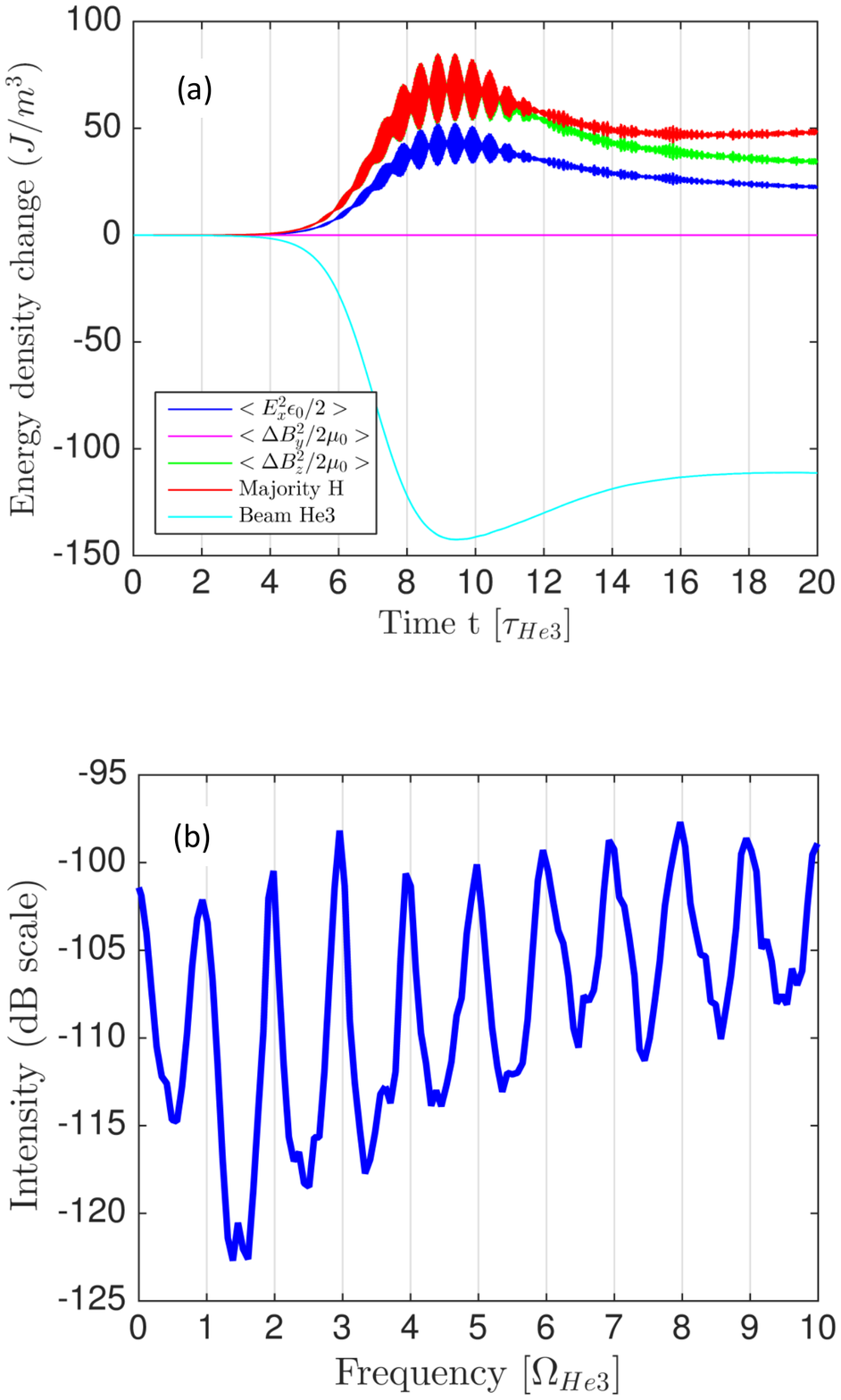}
\caption{\label{fig:power} (a) Changes in energy densities of particle species (bulk protons in red, energetic $^3$He ions in turquoise) and field components ($E_x$ in blue, $B_y$ in magenta, $B_z$ in green) during hybrid simulation with massless electrons, wavevector inclined at 90$^{\circ}$ with respect to $\bm{B}_0$ and a 10\% initial spread of velocities in the $^3$He distribution. Time is in units of the $^3$He cyclotron period. (b) Spectrum of fluctuations obtained by Fourier transforming $B_z$ over the entire space and time domain of the simulation.} 
\end{figure} 

\end{document}